\documentstyle[aps,12pt,pre,preprint]{revtex}
\tightenlines

\begin{document}
\draft
\title{Diffusion in a Granular Fluid - Theory}
\author{James W. Dufty}
\address{Department of Physics, University of Florida\\
Gainesville, FL 32611}
\author{J. Javier Brey}
\address{Fisica Te\'{o}rica, Universidad de Sevilla\\
Apartado de Correos 1065, E-41080 Sevilla, Spain}
\author{James Lutsko}
\address{Center for Nonlinear Phenomena and Complex Systems\\
Universit\'{e} Libre de Bruxelles\\
Campus Plaine, CP 231, 1050-Bruxelles, Belgium}
\date{\today }
\maketitle

\begin{abstract}
Many important properties of granular fluids can be represented by a system
of hard spheres with inelastic collisions. Traditional methods of
nonequilibrium statistical mechanics are effective for analysis and
description of the inelastic case as well. This is \ illustrated here for
diffusion of an impurity particle in a fluid undergoing homogeneous cooling.
An appropriate scaling of the Liouville equation is described such that the
homogeneous cooling ensemble and associated time correlation functions map
to those of a stationary state. In this form the familiar methods of linear
response can be applied, leading to Green - Kubo and Einstein
representations of diffusion in terms of the velocity and mean square
displacement correlation functions. These correlation functions are
evaluated approximately using a cumulant expansion and from kinetic theory,
providing the diffusion coefficient as a function of the density and the
restitution coefficients. Comparisons with results from molecular dynamics
simulation are given in the following companion paper.
\end{abstract}

\pacs{PACS number(s):45.70.Mg, 45.70.-n, 05.70.Ln}

\section{Introduction}

\label{sec1}

It has long been recognized that rapid flow granular media have properties
similar to those of ordinary fluids \cite{Ha83,Ca90}. Attempts to sharpen
this relationship have used idealized systems of hard spheres with inelastic
collisions. Remarkably, the single feature of inelasticity allows
reproduction of many qualitative features observed in real granular fluids.
This observation is based on the derivation of hydrodynamic equations from
kinetic theory \cite{LSJyCh84,JyR85,SyG98,BDKyS98,GyD99}, direct Monte Carlo
simulation of kinetic equations \cite{BRyC96,BRyM98}, molecular dynamics
simulation of dense fluids \cite{He95}, and controlled experiments on
inelastic hard spheres \cite{ZyS91,NHyT95,KWyG97,WHHPyA00,candela,rericha01}%
. Consequently, there is a growing opinion that the traditional methods of
nonequilibrium statistical mechanics applied to such model systems provide
the means to understand granular media at the most fundamental microscopic
level. The objective here is to give further support for this opinion by a
detailed application and test of linear response methods applied to
diffusion of an impurity particle in a fluid of inelastic hard spheres. Some
adaptation is required since the reference states are inherently
nonequilibrium, but the central ideas of linear response for normal fluids
are retained. A preliminary report of some of these results presented here
has been given in \cite{Du00}.

Diffusion is the prototype transport process and the associated diffusion
equation is the prototype hydrodynamic description for macroscopic dynamics.
For normal fluids, diffusion in a system of hard elastic spheres also has
been the testing ground for many body methods in nonequilibrium statistical
mechanics (density dependence, correlated many-body collisions, mode
coupling, percolation, glass transition). The benchmarks have been set by
accurate molecular dynamics simulation for normal fluids \cite{AAyY83} and
fluid-like lattice gas cellular automata \cite{NEHyF91}. In this way, the
conditions for macroscopic diffusion and the accuracy of methods for
predicting the diffusion coefficient are known over a wide range of
densities. There is evidence based on kinetic theory and molecular dynamics
simulation that similar studies are relevant for the system of inelastic
hard spheres \cite{BRCyG00,ByC01,NyE01}. On the other hand, there are
significant differences to confront. An isolated fluid of inelastic hard
spheres does not support a Gibbs state or any other stationary state, since
the collisions lead to a continual loss of energy or ``cooling''. Instead,
the analogue of the equilibrium state is a ''homogeneous cooling state''
(HCS) whose time dependence is entirely given via the mean square kinetic
energy of the particles \cite{GyS95,BDyS97,BRyC96,vanNoijehcs98}. As shown
below, the scaling property associated with this state allows a change of
variables in terms of which a stationary, but non-Gibbs, state results.

In the next section, the Liouville dynamics for inelastic hard spheres is
reviewed \cite{Du00,BDyS97,NyE01}. The corresponding nonequilibrium
statistical mechanics is given in terms of the Liouville equation for the
ensemble, and also in terms of the BBGKY hierarchy for the associated
reduced distribution functions. Next, it is shown that the Liouville
equation supports a scaling solution describing the HCS. A time dependent
temperature is defined in terms of the mean square velocity, in the same way
as for elastic collisions, hence the terminology ``homogeneous cooling
state''. A transformation to dimensionless variables allows a representation
of expectation values and time correlation functions in terms of stationary
state averages, just as for the Gibbs state.

The probability density for the position of a tagged or impurity particle as
a function of time is considered in Sec.\ \ref{sec3}. Linear response
methods (now inherently nonequilibrium linear response) are applied to
obtain a diffusion-like equation for long wavelengths, with a time dependent
diffusion coefficient. The analysis parallels closely that for fluids with
elastic collisions, except that the dimensionless time (the collision
number) required to accommodate the cooling is logarithmically stretched
compared to real time. Consequently, the mean square displacement approaches
linearity in the collision number rather than real time. The diffusion
coefficient is expressed as a time integral of the velocity autocorrelation
function which becomes a Green-Kubo expression for long times. This limit
also establishes the time scale for the validity of the diffusion equation,
and represents a clear example of macroscopic hydrodynamics for a granular
system.

An approximate evaluation of the velocity correlation function of the tagged
particle is carried out in Sec.\ \ref{sec4} by two different methods. The
first is based on a leading cumulant expansion, while the second is an
evaluation by means of kinetic theory. The two results are quite similar,
and their relationship is clearly identified. Also, it is confirmed in
Appendix \ref{ap3} that the evaluation of the Green-Kubo expression by
kinetic theory agrees with that obtained by the Chapman-Enskog solution to
the Enskog kinetic equation for the distribution function. Moreover, for
elastic collisions these approximations are known to give an accurate
description of the diffusion coefficient over a wide range of densities.
This accuracy cannot be assumed {\em a priori} to apply for inelastic
collisions, since the stationary state is nonequilibrium. The diffusion
coefficient is given as a function of the density, restitution coefficient,
and ratio of temperatures for the fluid and impurity particles. This latter
parameter is a peculiarity of granular fluids for mixtures, where the lack
of detailed balance leads to a HCS with all species having the same cooling
rate but different temperatures \cite{GyD99b}. The consequences for impurity
diffusion and mobility have been discussed elsewhere recently \cite%
{DyG01,SyD01}. The last section contains a summary and discussion of the
main results. The theoretical developments presented here are tested and
extended in a companion paper by comparison with molecular dynamics
simulation results over a wide range of densities and inelasticies for the
particular case of self diffusion \cite{LDyB01}.

\section{Statistical Mechanics and the Homogeneous Cooling State}

\label{sec2} The system considered is composed of a fluid of $N$ identical
hard disks or spheres (mass $m$, diameter $\sigma $, and fluid-fluid
coefficient of normal restitution $\alpha $), and an additional impurity
particle (mass $m_{0}$, diameter $\sigma _{0}$, and fluid-impurity
coefficient of normal restitution $\alpha _{0}$). The position and velocity
coordinates of the fluid particles will be denoted by $\{ {\bf q}_{i},{\bf v}%
_{i}; i=1, \ldots N \}$, while those for the impurity particle are ${\bf q}%
_{0},{\bf v}_{0}$. The dynamics consists of free streaming until a given
pair of fluid particles, $i,j$, or a fluid and impurity pair, $0,i$, is in
contact. At that point, the relative velocity of the pair changes
instantaneously according to the inelastic collision rules 
\begin{equation}  \label{2.1}
\widetilde{{\bf g}}_{ij}={\bf g}_{ij}-\left( 1+\alpha \right) \left( 
\widehat{\mbox{\boldmath $\sigma$}}\cdot {\bf g}_{ij}\right) \widehat{%
\mbox{\boldmath $\sigma$}}, \quad \widetilde{{\bf g}}_{0i}={\bf g}%
_{0i}-\left( 1+\alpha_{0}\right) \left( \widehat{\mbox{\boldmath $\sigma$}}%
\cdot {\bf g}_{0i}\right) \widehat{\mbox{\boldmath $\sigma$}}.
\end{equation}
Here ${\bf g}_{ij}={\bf v}_{i}-{\bf v}_{j}$ and ${\bf g}_{0i}={\bf v}_{0}-%
{\bf v}_{i}$ are the relative velocities, and $\bbox{\widehat{\sigma}}$ is a
unit vector directed from the center of particle $j$ to the center of
particle $i$ ($i$ to $0$, respectively) through the point of contact. The
coefficients of restitution have values in the range $0<\alpha,\alpha
_{0}\leq 1$, measuring the degree of inelasticity. The special case of
elastic particles is given by $\alpha =\alpha_{0}=1$. The center of mass
velocity is unchanged so that the total mass and momentum of the pairs are
conserved in such collisions. However, there is an energy loss for each
fluid-fluid particle collision 
\begin{equation}
\widetilde{E}_{ij}-E_{ij}=-\frac{m}{4}(1-\alpha ^{2}) (\widehat{%
\mbox{\boldmath $\sigma $}} \cdot {\bf g}_{ij})^{2}
\end{equation}
and for each fluid-impurity particle collision 
\begin{equation}
\widetilde{E}_{0i}-E_{0i}=-\frac{\mu}{2}(1-\alpha _{0}^{2}) (\widehat{%
\mbox{\boldmath $\sigma $}}\cdot{\bf g}_{0i})^{2},
\end{equation}
where $\mu =m_{0}m/\left( m_{0}+m\right)$. The state of the system at time $%
t $ is completely characterized by the positions and velocities of all
particles at that time and is represented by a point $\Gamma _{t}\equiv
\left\{ {\bf q}_{0}(t),\ldots,{\bf q}_{N}(t), {\bf v}_{0}(t),\dots,{\bf v}%
_{N}(t) \right\}$ in the associated $2d\left(N+1\right)$ dimensional phase
space, where $d=2$ for hard disks and $d=3$ for hard spheres. The sequence
of free streaming and binary collisions determines uniquely the positions
and velocities of the hard particles at time $t$ for given initial
conditions at $t^{\prime}<t$. A more complete notation expressing this
dependence on initial conditions is $\Gamma _{t}(\Gamma _{t^{\prime }})$.
Thus, just as in the case of elastic collisions, the microdynamics for this
system corresponds to a deterministic trajectory in phase space.

Observables of interest are represented by the same phase functions as for
elastic collisions, $A(\Gamma_{t})$, and their average for given statistical
initial data at $t=0$ is defined by 
\begin{equation}
\langle A(t);0\rangle \equiv \int d\Gamma \rho (\Gamma )A \left[
\Gamma_{t}(\Gamma) \right],  \label{2.2}
\end{equation}
where $\rho (\Gamma)$ is the probability density or ensemble for the initial
state, normalized to unity. An equivalent representation of this average is
obtained by changing variables to integrate over $\Gamma _{t}$ rather than
over $\Gamma $. This change of variables is possible since trajectories in
phase space do not cross, and $\Gamma$ can be expressed in terms of $\Gamma
_{t}$ denoted by $\Gamma _{t}^{-1}\left( \Gamma_{t}\right) $. This allows
the time dependence in Eq.\ (\ref{2.2}) to be expressed in terms of the
probability density 
\begin{equation}
\langle A(t);0 \rangle = \int d\Gamma\, \rho (\Gamma ,t)A(\Gamma ) \equiv
\langle A;t \rangle ,  \label{2.3}
\end{equation}
with the probability density at time $t$ given by 
\begin{equation}
\rho ( \Gamma ,t)\equiv J\left[ \Gamma _{t}^{-1}\left( \Gamma \right) ,
\Gamma \right] \rho \left[ \Gamma _{t}^{-1}\left( \Gamma \right) \right],
\label{2.4}
\end{equation}
$J\left( \Gamma,\Gamma_{t}\right)$ being the Jacobian of the transformation.

\subsection{Liouville Dynamics}

For practical purposes it is useful to identify the generators $L$ and $%
\overline{L}$ for the two above representations, defined by 
\begin{equation}
\langle A(t);0\rangle =\int d\Gamma \,\rho (\Gamma )e^{tL}A(\Gamma )=\int
d\Gamma \,\left[ e^{-t\overline{L}}\rho (\Gamma )\right] A(\Gamma ).
\label{2.6}
\end{equation}%
The last equality is consistent with the adjoint relationship implied by
Eqs.\ (\ref{2.2}) and (\ref{2.3}). These are not the usual generators of
Hamilton's equations for continuous forces, but are somewhat more complex
due to the singular nature of hard particles. Such generators have been
discussed in detail for the case of elastic collisions and the analysis
extends quite naturally to the inelastic case as well \cite%
{BDyS97,vNyE98,Du00,NyE01}, with the results 
\begin{equation}
L=L_{f}+{\bf v}_{0}\cdot {\bf \nabla }_{0}+\sum_{i=1}^{N}T(0,i),\quad 
\overline{L}=\overline{L}_{f}+{\bf v}_{0}\cdot {\bf \nabla }%
_{0}-\sum_{i=1}^{N}\overline{T}(0,i).  \label{2.7}
\end{equation}%
Here $L_{f}$ and $\overline{L}_{f}$ are the generators for the fluid
particles alone 
\begin{equation}
L_{f}=\sum_{i=1}^{N}{\bf v}_{i}\cdot {\bf \nabla }_{i}+\frac{1}{2}%
\sum_{i=1}^{N}\sum_{j\neq i}^{N}T(i,j),\quad \overline{L}_{f}=\sum_{i=1}^{N}%
{\bf v}_{i}\cdot {\bf \nabla }_{i}-\frac{1}{2}\sum_{i=1}^{N}\sum_{j\neq
i}^{N}\overline{T}(i,j).  \label{2.8}
\end{equation}%
The terms involving spatial gradients generate free streaming while the
others describe velocity changes. The binary collision operators $T(i,j)$
and $\overline{T}(i,j)$ for particles $i$ and $j$ are 
\begin{equation}
T(i,j)=\sigma ^{d-1}\int \ d\Omega \,\Theta (-{\bf g}_{ij}\cdot \widehat{%
\mbox{\boldmath $\sigma$}})|{\bf g}_{ij}\cdot \widehat{%
\mbox{\boldmath
$\sigma$}}|\delta ({\bf q}_{ij}-\mbox{\boldmath $\sigma$})(b_{ij}-1),
\label{2.9}
\end{equation}%
\begin{equation}
\overline{T}(i,j)=\sigma ^{d-1}\int d\Omega \,\Theta ({\bf g}_{ij}\cdot 
\widehat{\mbox{\boldmath $\sigma$}})|{\bf g}_{ij}\cdot \widehat{%
\mbox{\boldmath $\sigma$}}|[\alpha ^{-2}\delta ({\bf q}_{ij}-%
\mbox{\boldmath
$\sigma $})b_{ij}^{-1}-\delta ({\bf q}_{ij}+\mbox{\boldmath
$\sigma$})],  \label{2.10}
\end{equation}%
where $d\Omega $ denotes the solid angle element for the unit vector $%
\widehat{\mbox{\boldmath $\sigma$}}$, $\mbox{\boldmath $\sigma$}=\sigma 
\widehat{\mbox{\boldmath $\sigma$}}$, ${\bf q}_{ij}$ is the relative
position vector of the two particles, $\Theta $ is the Heaviside step
function, and $b_{ij}$ is a substitution operator, $b_{ij}F({\bf g}_{ij})=F(%
\widetilde{{\bf g}}_{ij})$, which changes the relative velocity ${\bf g}%
_{ij} $ into its scattered value $\widetilde{{\bf g}}_{ij}$, given by Eq.\ (%
\ref{2.1}). On the other hand, it does not change the velocity of the center
of mass of the two particles. The operator $b_{ij}^{-1}$ is the inverse of $%
b_{ij}$ and characterizes the ``restituting'' collision. The binary
operators for collisions between fluid particles and the impurity are
similar to those for collisions among fluid particles, 
\begin{equation}
T(0,i)=\overline{\sigma }^{d-1}\int d\Omega \,\Theta (-{\bf g}_{0i}\cdot 
\widehat{\mbox{\boldmath $\sigma$}})|{\bf g}_{0i}\cdot \widehat{%
\mbox{\boldmath $\sigma$}}|\delta ({\bf q}_{0i}-\overline{%
\mbox{\boldmath
$\sigma$}})\left( b_{0i}-1\right) ,  \label{2.11}
\end{equation}%
\begin{equation}
\overline{T}(0,i)=\overline{\sigma }^{d-1}\int d\Omega \,\Theta ({\bf g}%
_{i0}\cdot \widehat{\mbox{\boldmath $\sigma$}})|{\bf g}_{0i}\cdot \widehat{%
\mbox{\boldmath $\sigma$}}|[\alpha _{0}^{-2}\delta ({\bf q}_{0i}-\overline{%
\mbox{\boldmath $\sigma$}})b_{0i}^{-1}-\delta ({\bf q}_{0i}+\overline{%
\mbox{\boldmath $\sigma$}})],  \label{2.12}
\end{equation}%
with $\overline{\sigma }=\left( \sigma +\sigma _{0}\right) /2$ and $%
\overline{\mbox{\boldmath $\sigma$}}=\overline{\sigma }\widehat{%
\mbox{\boldmath $\sigma$}}$. The explicit forms for $b_{ij}^{-1}$ and $%
b_{0i}^{-1}$ are 
\begin{equation}
b_{ij}^{-1}{\bf g}_{ij}={\bf g}_{ij}-\frac{1+\alpha }{\alpha }\left( 
\widehat{\mbox{\boldmath $\sigma$}}\cdot {\bf g}_{ij}\right) \widehat{%
\mbox{\boldmath $\sigma$}},  \label{2.12a}
\end{equation}%
\begin{equation}
b_{0i}^{-1}{\bf g}_{0i}={\bf g}_{0i}-\frac{1+\alpha _{0}}{\alpha _{0}}\left( 
\widehat{\mbox{\boldmath $\sigma$}}\cdot {\bf g}_{0i}\right) \widehat{%
\mbox{\boldmath $\sigma$}}.  \label{2.12b}
\end{equation}%
The dynamics for the phase functions and the equivalent Liouville equation
distribution function, $\rho (\Gamma ,t)$, follow directly from Eq.\ (\ref%
{2.6}): 
\begin{equation}
\left( \partial _{t}-L\right) A(\Gamma ,t)=0,  \label{2.12c}
\end{equation}%
\begin{equation}
\left( \partial _{t}+\overline{L}\right) \rho (\Gamma ,t)=0.  \label{2.13}
\end{equation}

The BBGKY hierarchy of equations for the reduced distribution functions is
obtained by partial integration of the Liouville equation over the position
and velocities of $N-l$ fluid particles, 
\begin{equation}
\left[ \partial _{t}+\overline{L}(x_{0},\ldots ,x_{l})\right]
f^{(l+1)}(x_{0},\ldots ,x_{l};t)=\sum_{i=0}^{l}\int dx_{l+1}\,\overline{T}
(i,l+1)f^{(l+2)}(x_{0},\ldots ,x_{l+1};t),  \label{2.14}
\end{equation}
with the reduced distribution functions defined by 
\begin{equation}
f^{(l+1)}(x_{0},\ldots ,x_{l};t)\equiv N^{l}\int dx_{l+1}\ldots dx_{N}\,\rho
(\{x_{i} \};t),  \label{2.15}
\end{equation}
where $x_{i}\equiv \left\{ {\bf q}_{i},{\bf v}_{i} \right\}$ denotes the
position and velocity for particle $i$ and $\overline{L}(x_{0},\ldots,x_{l}) 
$ is the Liouville operator for a system of $l$ fluid particles and the
impurity particle. Moreover, the limit of large $N$ has been considered. The
above results in this section provide the basic tools and definitions of
nonequilibrium statistical mechanics for granular media \cite{BDyS97}.

\subsection{Homogeneous Cooling State (HCS)}

Stationary solutions to the Liouville equation (\ref{2.13}) are expected
when suitable external forces or boundary conditions are imposed. However,
there is no stationary solution for an isolated system, corresponding to the
spatially homogeneous Gibbs state, due to the inherent time dependence
following from loss of energy in collisions. This can be seen by calculating
the rate of change of the mean square velocity of a fluid particle in an
isolated state. For purposes below, the latter is used to define a kinetic
temperature according to 
\begin{equation}
T(t)\equiv \frac{1}{d}\left\langle m{\bf v}_{1}^{2};t\right\rangle \equiv 
\frac{1}{2}mv^{2}(t).  \label{2.17}
\end{equation}%
In addition to $T(t)$, Eq. (\ref{2.17}) defines the associated thermal
velocity $v(t)$ (a factor of Boltzmann's constant usual in elastic systems
has been deleted since there is no zeroth law of thermodynamics for granular
media; alternatively, to incorportate the elastic limit $\alpha =1$ the
temperature should understood as defined in units such that $k_{B}=1$). The
time dependence of these quantities can be calculated using the Liouville
dynamics to get 
\begin{equation}
\partial _{t}T(t)=-\zeta (t)T(t),  \label{2.18}
\end{equation}%
where $\zeta (t)$ is the ``cooling'' rate due to inelastic collisions, 
\begin{equation}
\zeta \left( t\right) =(1-\alpha ^{2})\frac{\sigma ^{d-1}}{2dNv^{2}(t)}\int d%
{\bf q}_{1}\int d{\bf v}_{1}\,\int d{\bf v}_{2}\,\int d\Omega \,\Theta ({\bf %
g}_{12}{\bf \cdot }\widehat{\mbox{\boldmath $\sigma $}})({\bf g}_{12}{\bf %
\cdot }\widehat{\mbox{\boldmath $\sigma $}})^{3}f^{(2)}({\bf q}_{1},{\bf v}%
_{1},{\bf q_{1}+\mbox{\boldmath $\sigma$}},{\bf v}_{2};t),  \label{2.19}
\end{equation}%
$f^{(2)}({\bf q}_{1},{\bf v}_{1},{\bf q_{1}+\mbox{\boldmath $\sigma$}},{\bf v%
}_{2};t)$ being the reduced two fluid particle distribution function at
contact. The latter is in general defined by 
\begin{equation}
f^{(2)}(x_{1},x_{2};t)=\int dx_{0}\,f^{(3)}(x_{0},x_{1},x_{2};t).
\label{2.19a}
\end{equation}%
For an homogeneous system, its spatial dependence occurs only through ${\bf q%
}_{12}$. Upon deriving Eq.\ (\ref{2.19}) we have taken into account that the
contribution from the impurity particle is negligible in the limit of large $%
N$.

In place of the Gibbs distribution, it is assumed that there is a
homogeneous scaling solution $\rho _{hcs}(\Gamma ,t)$ to the Liouville
equation, for which all time dependence occurs through a scaling of the
velocity (``cooling'') with the thermal velocity $v(t)$ \cite{GyS95,BDyS97}, 
\begin{equation}
\rho _{hcs}(\Gamma ,t)=\left[ \ell v(t)\right] ^{-d\left( N+1\right) }\rho
_{hcs}^{\ast }\left( \{{\bf q}_{ij}/\ell ,{\bf v}_{i}/v(t)\}\right) .
\label{2.16}
\end{equation}%
The dimensionless distribution function $\rho _{hcs}^{\ast }$ is invariant
under space translations, with the coordinates scaled (arbitrarily) relative
to $\ell \equiv (n\sigma ^{d-1})^{-1}$, which is proportional to the mean
free path ($n$ is the number density of particles). Therefore, $\rho
_{hcs}(\Gamma ,t)$ represents a spatially homogeneous fluid. Substitution of
Eq.\ (\ref{2.16}) into the Liouville equation gives 
\begin{equation}
\frac{1}{2}\zeta (t)\sum_{i=0}^{N}\frac{\partial }{\partial {\bf v}_{i}}%
\cdot ({\bf v}_{i}\rho _{hcs})+\overline{L}\rho _{hcs}=0.  \label{2.20}
\end{equation}%
The self-consistent solution to the coupled set of equations (\ref{2.18})
and (\ref{2.20}) determines the {\em homogeneous cooling state} (HCS). It is
the analogue of the Gibbs state for elastic collisions and reduces to it for 
$\alpha =1$. For $\alpha <1$, the exact solution is not known (it is not
simply a Gaussian in the velocities as for the Gibbs state) but its
existence is supported by results from Monte Carlo \cite{BRyC96} and
molecular dynamics simulations \cite{Huth00}.

The notation in Eq.\ (\ref{2.16}) does not make explicit the dependence of $%
\rho_{hcs}$ on parameters specific to the impurity particle. Since all
velocities have been scaled relative to the thermal velocity determined by
the fluid, there is a clear dependence on the mass ratio $m/m_{0}$, expected
from equipartition of energy. However, equipartition is strictly a property
of the equilibrium Gibbs state and does not apply for the HCS \cite{GyD99b}.
Consequently, in addition to a dependence on the mass ratio, there is also a
dependence on the temperature ratio $T/T_{0}$, where $T_{0}$ is the
temperature parameter for the impurity. The detailed form of the
relationship between $T_{0}(t)$ and $T(t)$ will be discussed later on, but
it follows from the condition that the cooling rates of the fluid particles
and the impurity particles are the same in the HCS, implying that $T/T_{0}$
is time independent. Through this section, the dependence of $\rho_{hcs}$ on
time independent parameters of the impurity particle will continue to be
suppressed, although it will become important in the subsequent discussion
of impurity diffusion.

Some interesting consequences follow from the velocity scaling of the
distribution function associated to the HCS. The reduced distribution
functions also have this property, so it is easily verified from Eq.\ (\ref%
{2.19}) that $\zeta \left( t\right) \propto T^{1/2}(t)$. Then, Eq.\ (\ref%
{2.18}) can be integrated for the explicit time dependence of $T(t)$, 
\begin{equation}
T(t)=T(t^{\prime })\left[ 1+\frac{1}{2}\zeta (t^{\prime })(t-t^{\prime })%
\right] ^{-2}.  \label{2.21}
\end{equation}%
The temperature is seen to have an algebraic decay in\ real time (Haff's law %
\cite{Ha83}). For the analysis of the HCS, it is more convenient to use the
dimensionless time scale 
\begin{equation}
s(t,t^{\prime })\equiv \int_{t^{\prime }}^{t}d\tau \nu _{c}(\tau ),
\label{2.22}
\end{equation}%
where $\nu _{c}(t)$ is an average collision frequency given by 
\begin{equation}
\nu _{c}\left( t\right) =v\left( t\right) /\ell .  \label{2.23}
\end{equation}%
Thus $s(t,t^{\prime })$ is a measure of the average number of collisions per
particle in the interval $(t^{\prime },t)$. The integral in Eq.\ (\ref{2.22}%
) can be performed using Eq.\ (\ref{2.21}) with the result: 
\begin{equation}
s(t,t^{\prime })=\frac{2}{\zeta ^{\ast }}\ln \left[ 1+\frac{1}{2}\zeta
(t^{\prime })\left( t-t^{\prime }\right) \right] ,  \label{2.24}
\end{equation}%
where we have introduced the dimensionless cooling rate 
\begin{equation}
\zeta ^{\ast }=\ell \zeta (t)/v(t).  \label{2.24a}
\end{equation}%
It follows from dimensional analysisis that $\zeta ^{\ast }$ is time
independent. The cooling in terms of the dimensionless time $s$ is
exponential, 
\begin{equation}
T(t)=T(t^{\prime })e^{-\zeta ^{\ast }s(t,t^{\prime })}  \label{2.25}
\end{equation}%
and, consequently, 
\begin{equation}
v(t)=v(t^{\prime })e^{-\zeta ^{\ast }s(t,t^{\prime })/2}.  \label{2.25a}
\end{equation}%
Since $\zeta $ is proportional to $(1-\alpha ^{2})$, there is a crossover
from logarithmic to linear relationship between the two time scales for weak
inelasticity.

Knowledge of the time dependence of $v(t)$ also implies that for many
average properties. For example, the average value of $A(\Gamma )$ in the
HCS can be written 
\begin{equation}
\langle A;t\rangle _{hcs}=\int d\Gamma \rho _{hcs}(\Gamma ,t)A(\Gamma )=\int
d\Gamma ^{\ast }\rho _{hcs}^{\ast }(\Gamma ^{\ast })A(\{\ell {\bf q}%
_{i}^{\ast },v(t){\bf v}_{i}^{\ast }\}).  \label{2.26}
\end{equation}%
Use has been made of the scaling form (\ref{2.16}) and $\Gamma ^{\ast }=\{%
{\bf q}_{i}^{\ast },{\bf v}_{i}^{\ast }\}=\{{\bf q}_{i}/\ell ,{\bf v}%
_{i}/v(t)\}$. This last result suggests that the transformation to
dimensionless form may admit a stationary state representation for the HCS.
To see that this is the case, define for a general state $\rho (\Gamma,t)$ 
\begin{equation}
\rho (\Gamma ,t)=\left[ \ell v(t)\right] ^{-d\left( N+1\right) }\rho ^{\ast
}\left( \Gamma ^{\ast },s \right),  \label{2.26a}
\end{equation}%
where we have introduced the same scaling for space, time, and velocity as
above. Substitution of this into Eq.\ (\ref{2.13}) gives the dimensionless
Liouville equation%
\begin{equation}
\left( \partial _{s}+\overline{{\cal L}}^{\ast }\right) \rho ^{\ast }(\Gamma
^{\ast },s)=0,  \label{2.26b}
\end{equation}%
with the definitions 
\begin{equation}
\overline{{\cal L}}^{\ast }=\frac{1}{2}\zeta ^{\ast }\left[ {\cal K}^{\ast
}+d(N+1)\right] +\overline{L}^{\ast },  \label{2.26d}
\end{equation}%
\begin{equation}
\overline{L}^{\ast }=\frac{1}{\nu _{c}(t)}\overline{L}=\left[ \overline{L}%
\right] _{\{{\bf q}_{i}={\bf q}_{i}^{\ast },{\bf v}_{i}={\bf v}_{i}^{\ast
}\}},  \label{2.26c}
\end{equation}%
\begin{equation}
{\cal K}^{\ast }=\sum_{i=0}^{N}{\bf v}_{i}^{\ast }\cdot \frac{\partial }{%
\partial {\bf v}_{i}^{\ast }}.  \label{2.26c1}
\end{equation}%
This transformation of the Liouville equation explicitly accounts for the
collisional cooling, and in this form stationary solutions are now possible.
In fact (\ref{2.20}) becomes 
\begin{equation}
\overline{{\cal L}}^{\ast }\rho _{hcs}^{\ast }=0,  \label{2.26e2}
\end{equation}%
so the dimensionless HCS is a stationary solution to (\ref{2.26b}).

The average value of $A(\Gamma )$ for a general state $\rho (\Gamma ,t)$
becomes 
\begin{eqnarray}
\langle A;t\rangle &=&\int d\Gamma \rho (\Gamma ,t)A(\Gamma )=\int d\Gamma
^{\ast }\rho ^{\ast }(\Gamma ^{\ast },s)A(\{\ell {\bf q}_{i}^{\ast },v(t)%
{\bf v}_{i}^{\ast }\})  \nonumber \\
&=&\int d\Gamma ^{\ast }\left[ \bar{U}(s,0)\rho ^{\ast }(\Gamma ^{\ast },0)%
\right] A(\{\ell {\bf q}_{i}^{\ast },v(t){\bf v}_{i}^{\ast }\})  \nonumber \\
&\equiv &\int d\Gamma ^{\ast }\rho ^{\ast }(\Gamma ^{\ast },0)U(s,0)A(\{\ell 
{\bf q}_{i}^{\ast },v(t){\bf v}_{i}^{\ast }\}),  \label{2.26e3}
\end{eqnarray}%
where $\bar{U}(s,s^{\prime })$ and $U(s,s^{\prime })$ obey the equations 
\begin{equation}
\left( \partial _{s}+\bar{{\cal L}}^{\ast }\right) \bar{U}(s,s^{\prime
})=0,\quad \left( \partial _{s}-{\cal L}^{\ast }\right) U(s,s^{\prime })=0,
\label{2.26aa}
\end{equation}%
with the initial conditions $\bar{U}(s^{\prime },s^{\prime })=U(s^{\prime
},s^{\prime })=1$. The new generator for the dynamics of the phase functions
is 
\begin{equation}
{\cal L}^{\ast }=\frac{1}{2}\zeta ^{\ast }{\cal K}^{\ast }+L^{\ast },\quad
L^{\ast }=\left[ L\right] _{\{{\bf q}_{i}={\bf q}_{i}^{\ast },{\bf v}_{i}=%
{\bf v}_{i}^{\ast }\}}.  \label{2.26e4}
\end{equation}%
For the special case $\rho ^{\ast }(\Gamma ^{\ast },0)=\rho _{hcs}^{\ast
}(\Gamma ^{\ast })$, Eq. (\ref{2.26e3}) reduces to (\ref{2.26}).

The stationary representation is the most natural one for both theoretical
developments and for computer simulation, as is discussed in the following
companion paper \cite{LDyB01}. Similarly, the physically relevant time
scales are those expressed in terms of the average collision number $s$
rather than the real time $t$. It is appropriate at this point to note that
although $\rho _{hcs}^{\ast }$ is a stationary solution to (\ref{2.26b}),
there is convincing evidence from both theory and simulation that it is
unstable to long wavelength spatial perturbations and spontaneous
fluctuations \cite{GyZ93,McyY94}. In the following sections, time
correlation functions are considered for the HCS and use is made of
stationarity and spatial homogeneity. The results must be understood as
applying to system sizes for which the instability does not occur, or on
time scales that are short compared to those required for growth of spatial
structures.

\subsection{HCS Averages and Correlation Functions}

The HCS time correlation function for two phase functions $A(\Gamma )$ and $%
B(\Gamma )$ is defined as 
\begin{equation}
C_{AB}(t,t^{\prime })\equiv \langle A(t)B(t^{\prime });0\rangle -\langle
A(t);0\rangle \langle B(t^{\prime });0\rangle ,  \label{2.27}
\end{equation}%
with $t\geq t^{\prime }\geq 0$. Here and below the brackets $\langle
;t\rangle $ denote an average over the HCS at time $t$. For a system with
elastic collisions in equilibrium, the above autocorrelation function can be
reduced to a single time correlation function, using time translational
invariance and stationarity of the Gibbs state. In the case of inelastic
particles, the HCS is not stationary, but the scaling property (\ref{2.16})
can be used to transform the correlation function to an effective time
stationary average. First, use time translational invariance to write 
\begin{equation}
\langle A(t)B(t^{\prime });0\rangle =\langle A(t-t^{\prime })B(0);t^{\prime
}\rangle .  \label{2.28}
\end{equation}%
Next, transform to dimensionless variables to get%
\begin{eqnarray}
\langle A(t)B(t^{\prime });0\rangle &=&\int d\Gamma \rho _{hcs}(\Gamma
,t^{\prime })A(t-t^{\prime })B(0)  \nonumber \\
&=&\int d\Gamma ^{\ast }\rho _{hcs}^{\ast }(\Gamma ^{\ast })\left[ e^{\nu
_{c}(t^{\prime })(t-t^{\prime })L^{\ast }}A(\{\ell {\bf q}_{i}^{\ast
},v(t^{\prime }){\bf v}_{i}^{\ast }\})\right] B(\{\ell {\bf q}_{i}^{\ast
},v(t^{\prime }){\bf v}_{i}^{\ast }\}),  \label{2.29}
\end{eqnarray}%
where (\ref{2.26}) has been used. Next note the identity%
\begin{equation}
e^{-\frac{1}{2}\zeta ^{\ast }s(t,t^{\prime }){\cal K}^{\ast }}F\left( \{{\bf %
v}_{i}^{\ast }\right\} )=F\left( \left\{ e^{-\frac{1}{2}\zeta ^{\ast
}s(t,t^{\prime })}{\bf v}_{i}^{\ast }\right\} \right) =F\left( \left\{ \frac{%
v(t)}{v(t^{\prime })}{\bf v}_{i}^{\ast }\right\} \right) .  \label{2.29a}
\end{equation}%
The correlation function now can be written%
\begin{equation}
\langle A(t)B(t^{\prime });0\rangle =\int d\Gamma ^{\ast }\rho _{hcs}^{\ast
}(\Gamma ^{\ast })\left[ U(t,t^{\prime })A(\{\ell {\bf q}_{i}^{\ast },v(t)%
{\bf v}_{i}^{\ast }\})\right] B(\ell {\bf q}_{i}^{\ast },v(t^{\prime }){\bf v%
}_{i}^{\ast }\}),  \label{2.29b}
\end{equation}%
where 
\begin{equation}
U(t,t^{\prime })=e^{\nu _{c}(t^{\prime })(t-t^{\prime })L^{\ast }}e^{\frac{1%
}{2}\zeta ^{\ast }s(t,t^{\prime }){\cal K}^{\ast }}.  \label{2.29c}
\end{equation}%
This time evolution operator can be identified by differentiating with
respect to $s(t,t^{\prime })$, taking into account that in the HCS $\zeta
^{\ast }$ is time independent, 
\begin{eqnarray}
\frac{\partial U(t,t^{\prime })}{\partial s} &=&U(t,t^{\prime })\left[ \frac{%
\partial t}{\partial s}e^{-\frac{1}{2}\zeta ^{\ast }s(t,t^{\prime }){\cal K}%
^{\ast }}\nu _{c}(t^{\prime })L^{\ast }e^{\frac{1}{2}\zeta ^{\ast
}s(t,t^{\prime }){\cal K}^{\ast }}+\frac{1}{2}\zeta ^{\ast }{\cal K}^{\ast }%
\right]  \nonumber \\
&=&U(t,t^{\prime })\left[ \frac{\nu _{c}(t^{\prime })}{\nu _{c}(t)}e^{-\frac{%
1}{2}\zeta ^{\ast }s(t,t^{\prime })}L^{\ast }+\frac{1}{2}\zeta ^{\ast }{\cal %
K}^{\ast }\right] =U(t,t^{\prime }){\cal L}^{\ast }.  \label{2.29d}
\end{eqnarray}%
Consequently,%
\begin{equation}
U(t,t^{\prime })=e^{s(t,t^{\prime }){\cal L}^{\ast }}  \label{2.29e}
\end{equation}%
Note that this is propagator is the same as that in Eq. \ref{2.26aa}
specialized to the HCS. In this case $\zeta ^{\ast }$ becomes time
independent, allowing the simple exponential representation. The correlation
function now can be written in the final form%
\begin{equation}
\langle A(t)B(t^{\prime });0\rangle =\int d\Gamma ^{\ast }\rho _{hcs}^{\ast
}(\Gamma ^{\ast })\left[ e^{s(t,t^{\prime }){\cal L}^{\ast }}A(\{\ell {\bf q}%
_{i}^{\ast },v(t){\bf v}_{i}^{\ast }\})\right] B(\ell {\bf q}_{i}^{\ast
},v(t^{\prime }){\bf v}_{i}^{\ast }\}).  \label{2.29f}
\end{equation}

This is a primary result of this section. The time correlation functions
depend on the dynamics through the collision number $s(t,t^{\prime })$. All
additional time dependence occurs trivially through the thermal velocity.
This is most evident when $A$ and $B$ are homogeneous functions of the
velocity,%
\begin{equation}
A(\{\ell {\bf q}_{i}^{\ast },v(t){\bf v}_{i}^{\ast }\})=v^{a}(t)A(\{\ell 
{\bf q}_{i}^{\ast },{\bf v}_{i}^{\ast }\}),\hspace{0.3in}B(\ell {\bf q}%
_{i}^{\ast },v(t^{\prime }){\bf v}_{i}^{\ast }\})=v^{b}(t^{\prime })B(\{\ell 
{\bf q}_{i}^{\ast },{\bf v}_{i}^{\ast }\}).  \label{2.30}
\end{equation}%
Then the correlation function becomes 
\begin{equation}
\langle A(t)B(t^{\prime });0\rangle =v^{a}(t)v^{b}(t^{\prime })\langle
A(s)B\rangle ^{\ast },  \label{2.31}
\end{equation}%
\begin{equation}
\langle A(s)B\rangle ^{\ast }=\int d\Gamma ^{\ast }\rho^{\ast}_{hcs}(\Gamma
^{\ast })A(\{\ell {\bf q}_{i}^{\ast },{\bf v}_{i}^{\ast }\},s)B(\{\ell {\bf q%
}_{i}^{\ast },{\bf v}_{i}^{\ast }\}),  \label{2.32}
\end{equation}%
and the phase function $A(\{\ell {\bf q}_{i}^{\ast },{\bf v}_{i}^{\ast
}\},s) $ is%
\begin{equation}
A(\{\ell {\bf q}_{i}^{\ast },{\bf v}_{i}^{\ast }\},s)=e^{s(t,t^{\prime })%
{\cal L}^{\ast }}A(\{\ell {\bf q}_{i}^{\ast },{\bf v}_{i}^{\ast }\})
\label{2.33}
\end{equation}%
This stationary state representation for the time correlation functions
simplifies considerably the analysis of response functions in the next
section.

\section{Impurity particle diffusion}

\label{sec3}

In this section, the diffusion equation and associated expressions for the
diffusion coefficient are derived for a granular system in the HCS. The
probability density $P({\bf r},t)$ to find the tagged or impurity particle
at point ${\bf r}$ at time $t$, given it was at the origin at $t=0$, is
defined by 
\begin{equation}
P({\bf r},t)=V \langle \delta [{\bf q}_{0}(t)-{\bf r}]\delta ({\bf q}_{0});0
\rangle =\langle \delta [{\bf q}_{0}(t)-{\bf q}_{0}-{\bf r}];0 \rangle,
\label{3.1}
\end{equation}
where the angular brackets indicate as above an average over an initial HCS
and $V$ is the volume (for $d=3$) or surface (for $d=2$) of the system. The
second equality is a consequence of the translational invariance of the HCS.
The conservation law for probability follows by differentiation of Eq.\ (\ref%
{3.1}) with respect to $t$, 
\begin{equation}
\partial _{t}P({\bf r},t)+\nabla \cdot {\bf J}({\bf r},t)=0,  \label{3.2}
\end{equation}
with the probability flux ${\bf J}({\bf r},t)$ identified as 
\begin{equation}
{\bf J}({\bf r},t)=\langle {\bf v}_{0}(t)\delta [{\bf q}_{0}(t)-{\bf q}_{0} -%
{\bf r}];0 \rangle.  \label{3.3}
\end{equation}

The interest here is in the limiting behavior of Eq.\ (\ref{3.2}) in the
hydrodynamic regime, which corresponds to the long wavelength region. The
long wavelength spatial dependence of ${\bf J}({\bf r},t)$ can be obtained
from a Fourier representation of $P({\bf r},t)$, 
\begin{equation}
\widetilde{P}({\bf k},t)=\int d{\bf r}\, e^{i{\bf k\cdot r}}P({\bf r},t).
\label{3.4}
\end{equation}
To get a formal equation for $\widetilde{P}({\bf k},t)$, it is useful to
introduce the index of the distribution $\widetilde{C}({\bf k},t)$ by 
\begin{equation}
\widetilde{P}({\bf k},t)=e^{\widetilde{C}({\bf k},t)},  \label{3.5}
\end{equation}
i.e., 
\begin{equation}  \label{3.5a}
\widetilde{C}({\bf k},t)=\ln \langle e^{i{\bf k\cdot }\left[ {\bf q}_{0}(t) -%
{\bf q}_{0}(0)\right] };0\rangle.
\end{equation}
Differentiation with respect to time of Eq.\ (\ref{3.5}) yields 
\begin{equation}
\left[\partial _{t}-\dot{\widetilde{C}}({\bf k},t)\right] \widetilde{P}({\bf %
k},t)=0,  \label{3.6}
\end{equation}%
where the dot over $\widetilde{C}$ denotes the derivative with respect to
time. For long wavelengths ($k\ell \ll 1$), $\dot{\widetilde{C}}({\bf k},t)$
can be expanded to order $k^{2}$, 
\begin{eqnarray}
\dot{\widetilde{C}}({\bf k},t) &=&\frac{\langle i{\bf k\cdot v}_{0}(t)e^{i%
{\bf k\cdot }\left[ {\bf q}_{0}(t)-{\bf q}_{0}(0)\right] };0 \rangle}{%
\langle e^{i{\bf k\cdot }\left[ {\bf q}_{0}(t)-{\bf q}_{0}(0)\right]}
;0\rangle}  \nonumber \\
&\simeq &-\frac{k^{2}}{d}\langle {\bf v}_{0}(t){\bf \cdot}\left[ {\bf q}%
_{0}(t)-{\bf q}_{0}(0)\right] ;0 \rangle.  \label{3.7}
\end{eqnarray}%
Substitution of this expression into Eq\ (\ref{3.6}) and inverting the
transform gives Eq.\ (\ref{3.3}) with the identification of the probability
flux as 
\begin{equation}
{\bf J}({\bf r},t)=-D(t)\nabla P({\bf r},t).  \label{3.8}
\end{equation}%
The finite time diffusion coefficient is 
\begin{equation}
D(t)=\frac{1}{2d}\frac{\partial}{\partial t}\, \langle \left| {\bf q}_{0}(t)
-{\bf q}_{0}(0)\right|^{2};0 \rangle.  \label{3.9}
\end{equation}
This will be referred to as the Einstein form, relating the diffusion
coefficient to the mean square displacement. The equivalent Green-Kubo form,
in terms of the velocity autocorrelation function (VACF), is derived by
using the relationship 
\begin{equation}
{\bf q}_{0}(t)-{\bf q}_{0}(0)=\int_{0}^{t}dt^{\prime }{\bf v}%
_{0}(t^{\prime}),  \label{3.10}
\end{equation}
with the result 
\begin{equation}
D(t)=\frac{1}{d}\int_{0}^{t}dt^{\prime } \langle {\bf v}_{0}(t)\cdot {\bf v}%
_{0}\left( t^{\prime }\right) ;0 \rangle.  \label{3.11}
\end{equation}

For normal fluids with elastic collisions, the diffusion {\em constant}
follows from the long time limit $D=\lim_{t\rightarrow \infty }D(t)$, or
equivalently from the coefficient of the mean square displacement when it
becomes linear in $t$. This limit occurs for times large compared to the
mean free time. Since the latter is time dependent in the HCS, the usual
conditions to establish a diffusion constant, and consequently the diffusion
equation, must be modified for granular media. This is done by introducing
the dimensionless diffusion coefficient 
\begin{equation}
D^{\ast }(s)=\frac{D(t)}{\ell ^{2}\nu _{c}\left( t\right) }\,.  \label{3.12}
\end{equation}%
Using the representation (\ref{2.31}) for the correlation function in (\ref%
{3.11}) gives the Green-Kubo form as a stationary \ average 
\begin{eqnarray}
D^{\ast }(s) &=&\frac{1}{d\ell ^{2}\nu _{c}\left( t\right) }%
\int_{0}^{t}dt^{\prime }v(t)v(t^{\prime })\langle {\bf v}_{0}^{\ast
}(s)\cdot {\bf v}_{0}^{\ast }\rangle ^{\ast }  \nonumber \\
&=&\frac{1}{d}\int_{0}^{t}dt^{\prime }v_{c}(t^{\prime })\langle {\bf v}%
_{0}^{\ast }(s-s^{\prime })\cdot {\bf v}_{0}^{\ast }\rangle ^{\ast } 
\nonumber \\
&=&\frac{1}{d}\int_{0}^{s}ds^{\prime }\langle {\bf v}_{0}^{\ast
}(s-s^{\prime })\cdot {\bf v}_{0}^{\ast }\rangle ^{\ast }  \nonumber \\
&=&\frac{1}{d}\int_{0}^{s}ds^{\prime }\langle {\bf v}_{0}^{\ast }(s^{\prime
})\cdot {\bf v}_{0}^{\ast }\rangle ^{\ast }.  \label{3.14}
\end{eqnarray}%
In going from the first line to the second line we have written $%
s(t,t^{\prime })=$ $s(t,0)-s(t^{\prime },0)\equiv s-s^{\prime }$. Similarly,
from Eqs.(\ref{3.9}) the corresponding Einstein form is 
\begin{equation}
D^{\ast }(s)=\frac{1}{2d}\frac{\partial }{\partial s}\,\langle \left| {\bf q}%
_{0}^{\ast }(s)-{\bf q}_{0}^{\ast }(0)\right| ^{2}\rangle ^{\ast }.
\label{3.13}
\end{equation}%
These are the stationary average representations for the diffusion
coefficient, and are the primary results of this section. In terms of the
time scale $s$, the mean square displacement is expected to become linear
for $s\gg 1$, and the velocity autocorrelation function is expected to decay
to zero also on this time scale. The physical interpretation of this limit
is the same as for elastic collisions, since $s$ is essentially the number
of collisions per particle. However, due to the time dependence of the
collision frequency, the correlation functions are expected to have the
proper behavior with respect to $s$ rather than $t$. This will be shown more
explicitly in the next section. The dimensionless form of (\ref{3.2}) and (%
\ref{3.8}) reads 
\begin{equation}
\partial _{s}P^{\ast }({\bf r}^{\ast },s)-D^{\ast }(s)\nabla ^{\ast
2}P^{\ast }({\bf r}^{\ast },s)=0,  \label{3.14a}
\end{equation}%
where $P^{\ast }({\bf r}^{\ast },s)=\ell ^{d}P({\bf r},t)$. Clearly, this
becomes the usual diffusion equation for sufficiently large $s$, if $D^{\ast
}(s)$ tends to a constant.

It is useful to introduce a dimensionless VACF for the impurity particle
that is normalized to unity at $s=0$. This requires calculation of $\langle
v_{0}^{\ast 2}\rangle^{\ast}$. It might appear from Eq.\ (\ref{2.17}) that
this can be obtained simply in terms of the mass of the impurity and the
temperature of the fluid. However, it has been shown elsewhere that
mechanically different particles in a common HCS do not have the same
temperatures \cite{GyD99b}, as already mentioned below Eq.\ (\ref{2.20}).
Thus $<v_{0}^{\ast 2}>^{\ast}$ is given by Eq.\ (\ref{2.17}) with both $m$
and $T(t)$ replaced by $m_{0}$ and $T_{0}(t)$, 
\begin{equation}
<v_{0}^{\ast 2}>^{\ast}=\frac{dT_{0}(t)}{m_{0}v^{2}(t)}=\frac{d}{2} \phi
_{hcs},  \label{3.15}
\end{equation}
where $\phi_{hcs}$ is the ratio of the square of the thermal velocity for
the impurity particle relative to that for the fluid particles, 
\begin{equation}
\phi _{hcs}=\frac{mT_{0}(t)}{m_{0}T(t)}.  \label{3.16}
\end{equation}%
Since the cooling rates of the fluid, $\zeta(t)$, and the impurity particle, 
$\zeta_{0}(t)$, are the same and they are proportional to the square root of
the temperature, the above ratio is time independent. The condition for
equal cooling rates also determines $\phi_{hcs} $. These rates are
calculated to good approximation using a local equilibrium ensemble in
Appendix \ref{ap2}, and are given by 
\begin{equation}
\zeta ^{\ast}=\frac{2^{1/2} \pi^{\frac{d-1}{2}}}{ \Gamma(d/2)d}\, \chi
(1-\alpha^{2}),  \label{3.17}
\end{equation}
\begin{equation}  \label{3.17a}
\zeta_{0}^{\ast}=\nu^{\ast} \left( 1- h \frac{1+\phi_{hcs}}{\phi_{hcs}}
\right) \left(1+\phi_{hcs} \right)^{1/2},
\end{equation}
where $h=\left( 1+\alpha _{0}\right) m/2\left( m+m_{0}\right) $, and $%
\nu^{\ast }$ is a dimensionless impurity particle collision rate, 
\begin{equation}
\nu ^{\ast }= \frac{4h}{d} \left( \frac{\overline{\sigma}}{\sigma}
\right)^{d-1} \frac{\pi^{\frac{d-1}{2}}}{\Gamma (d/2)}\, \chi_{0}.
\label{3.17b}
\end{equation}
The factors $\chi$ and $\chi_{0}$ are the pair correlation function for the
fluid-fluid and the fluid-impurity particles at contact, respectively. A
more accurate calculation for the case of hard spheres ($d=3$) is carried
out in Ref.\ \cite{DyG01}. Equating Eqs.\ (\ref{3.17}) and (\ref{3.17a})
provides the equation for $\phi _{hcs}$ 
\begin{equation}
\left( 1+\phi _{hcs}\right) ^{1/2}\left( 1-h\frac{1+\phi _{hcs}}{\phi _{hcs}}
\right) =\frac{\zeta ^{\ast }}{\nu ^{\ast }}\, .  \label{3.18}
\end{equation}
This gives a cubic equation which has a unique real, positive solution for
all allowed values of $h$ and $\zeta ^{\ast }/\nu ^{\ast }$. For elastic
collisions, $\phi _{hcs}\rightarrow m/m_{0}$ as required by the
equipartition theorem. Qualitative changes in this solution, similar to a
phase transition occur in the limit $h\rightarrow 0$ \cite{SyD01}, but will
not be discussed here. The normalized VACF is now given by 
\begin{equation}
C_{vv}^{\ast }(s)\equiv \frac{ \langle {\bf v}_{0}^{\ast }\left( s\right)
\cdot {\bf v}_{0}^{\ast }\rangle^{\ast}}{\langle v_{0}^{\ast 2}\rangle^{\ast}%
} =\frac{2}{d\phi _{hcs}}\langle{\bf v} _{0}^{\ast }\left( s\right) \cdot 
{\bf v}_{0}^{\ast }\rangle^{\ast}  \label{3.19}
\end{equation}
and the diffusion coefficient in Eq.\ (\ref{3.14}) becomes 
\begin{equation}
D^{\ast }(s)=\frac{\phi _{hcs}}{2} \int_{0}^{s}ds^{\prime }C_{vv}^{\ast}
(s^{\prime }).  \label{3.20}
\end{equation}

\section{Approximations}

\label{sec4} In the following, two approximations, originally developed for
fluids with elastic collisions \cite{McL89}, are applied to calculate the
VACF in the HCS. The first method uses a leading order truncation of a
cumulant expansion, while the second uses an approximate kinetic equation.
The results confirm the expected time scale for transition to hydrodynamics
and provide the detailed dependence on density and restitution coefficients.

\subsection{Cumulant expansion}

The cumulant expansion of the VACF is 
\begin{equation}
C_{vv}^{\ast }(s)=\exp \left[ \sum_{p=1}^{\infty }\frac{1}{p!}
\omega^{\ast}_{p}\left( -s\right) ^{p} \right] .  \label{4.1}
\end{equation}%
The coefficients $\omega _{p}^{\ast}$ are determined from the initial time
derivatives of the correlation function. Clearly, truncation of this
expansion at any order is asymptotically exact at short times and also for
small $h$ (heavy impurity) since each time derivative contributes a factor
of $h$. The simplest such approximation retains only the leading term 
\begin{equation}
C_{vv}^{\ast }(s)\simeq e^{-\omega _{1}^{\ast }s},  \label{4.2}
\end{equation}
with 
\begin{equation}
\omega _{1}^{\ast }= -\left[ \frac{\partial}{\partial s} \ln C_{vv} (s) %
\right]_{s=0} =-\frac{2}{d\phi _{hcs}}<\left( {\cal L}^{\ast }{\bf v}%
_{0}^{\ast }\right) \cdot {\bf v}_{0}^{\ast }>^{\ast }.  \label{4.3}
\end{equation}
Use has been made of the definition of the $s$ dependence in (\ref{3.19}).
The corresponding approximation for the mean square displacement is obtained
by integrating Eq.\ (\ref{4.2}) twice, 
\begin{equation}  \label{4.3a}
<| {\bf q}_{0}^{\ast }(s)-{\bf q}_{0}^{\ast }(0)|^{2}>^{\ast } = \frac{d\phi
_{hcs}}{\omega _{1}^{\ast }}\left[ s-\omega _{1}^{\ast -1}\left(
1-e^{-\omega _{1}^{\ast }s}\right) \right].
\end{equation}
For elastic collisions, this approximation is known to be accurate for short
as well as long times, and for a wide range of densities and mass ratios.
The resulting diffusion coefficient and diffusion constant are then found by
substituting Eq.\ (\ref{4.2}) into Eq.\, (\ref{3.20}), 
\begin{equation}
D^{\ast }(s)=D^{\ast }\left( 1-e^{-\omega _{1}^{\ast }s}\right),  \label{4.4}
\end{equation}
\begin{equation}
D^{\ast }=\lim_{s\rightarrow \infty }D^{\ast }(s)=\frac{\phi _{hcs}}{2\omega
_{1}^{\ast }}\, .  \label{4.5}
\end{equation}
Clearly $\omega _{1}^{\ast }$ is a characteristic dimensionless collision
frequency for the impurity particle. The first cumulant approximation
confirms the expectation that the mean square displacement becomes linear in 
$s$ and the velocity autocorrelation function decays for $s>>\omega
_{1}^{\ast }$. Consequently, the macroscopic diffusion equation applies on
this time scale as well. The collision frequency $\omega _{1}^{\ast }$ is
evaluated in Appendix \ref{ap2} by using a local equilibrium ensemble, with
the result: 
\begin{equation}
\omega _{1}^{\ast }=-\frac{1}{2}\zeta ^{\ast }+\nu ^{\ast } \left(
1+\phi_{hcs} \right)^{1/2},  \label{4.6}
\end{equation}%
where $\zeta ^{\ast }$and $\nu ^{\ast }$ are defined in (\ref{3.17}) and (%
\ref{3.17b}), respectively. Substitution of this into Eq.\ (\ref{4.5}) gives 
\begin{equation}
D^{\ast }=\frac{\phi _{hcs}}{2\left(1+\phi_{hcs}\right)^{1/2} \nu ^{\ast }
-\zeta ^{\ast } }.  \label{4.7}
\end{equation}
It is possible to show that $D^{\ast }$ is positive and finite for all
values of the density and restitution coefficients.

\subsection{Kinetic Theory}

Perhaps the most accurate and detailed evaluation of time correlation
functions is via kinetic theory methods. These can be applied as well to the
case of inelastic collisions \cite{vNyE98,BMyR98a,Du00}. To show this, first
use the adjoint property of the Liouville operators to write the velocity
autocorrelation function in the form 
\begin{eqnarray}
C_{vv}^{\ast }(s) &=&\frac{2}{d\phi _{hcs}}\int d\Gamma ^{\ast }\rho
_{hcs}^{\ast }(\Gamma ^{\ast }){\bf v}_{0}^{\ast }\cdot e^{s{\cal L}^{\ast }}%
{\bf v}_{0}^{\ast }  \nonumber \\
&=&\frac{2}{d\phi _{hcs}}\int d\Gamma ^{\ast }{\bf v}_{0}^{\ast }\cdot e^{-s%
\overline{{\cal L}}^{\ast }}\left[ \rho _{hcs}^{\ast }(\Gamma ^{\ast }){\bf v%
}_{0}^{\ast }\right] .  \label{4.8}
\end{eqnarray}%
Next, define the dimensionless reduced correlation functions in a manner
similar to (\ref{2.15}) 
\begin{equation}
\mbox{\boldmath  $\psi$}^{\ast (l+1)}(x_{0}^{\ast },\ldots ,x_{l}^{\ast
},s)\equiv N^{l}\int dx_{l+1}^{\ast }\ldots dx_{N}^{\ast }\,e^{-s\overline{%
{\cal L}}^{\ast }}\left[ \rho _{hcs}^{\ast }(\Gamma ^{\ast }){\bf v}%
_{0}^{\ast }\right] ,  \label{4.9}
\end{equation}%
so that the VACF can be written as 
\begin{equation}
C_{vv}^{\ast }(s)=\frac{2}{d\phi _{hcs}}\int dx_{0}^{\ast }\,{\bf v}%
_{0}^{\ast }\cdot \mbox{\boldmath  $\psi$ }^{\ast (1)}(x_{0}^{\ast },s).
\label{4.10}
\end{equation}%
It is easily verified by direct differentiation that the $%
\mbox{\boldmath
$\psi$}^{(l+1)}$ functions obey a hierarchy of equations similar to (\ref%
{2.14}), the first of which is 
\begin{equation}
\left( \partial _{s}+{\bf v}_{0}^{\ast }\cdot {\bf \nabla }_{0}^{\ast
}\right) \mbox{\boldmath $\psi$ }^{(1)}(x_{0}^{\ast },s)+\frac{1}{2}\zeta
^{\ast }\frac{\partial }{\partial {\bf v}_{o}^{\ast }}\cdot \left[ {\bf v}%
_{0}^{\ast }\mbox {\boldmath $\psi$}^{\ast (1)}(x_{0}^{\ast },s)\right]
=\int dx_{1}^{\ast }\,\overline{T}^{\ast }(0,1)\mbox{\boldmath $\psi$}^{\ast
(2)}(x_{0}^{\ast },x_{1}^{\ast },s).  \label{4.11}
\end{equation}

A kinetic equation results from a closure approximation in the above
equation that expresses $\mbox{\boldmath $\psi$}^{\ast (2)}$ as a functional
of $\mbox{\boldmath $\psi$}^{\ast (1)}$. Formally this is possible since
both $\mbox{\boldmath $\psi$}^{\ast (1)} (x_{0}^{\ast},s)$ and $%
\mbox{\boldmath $\psi$}^{\ast (2)}(x_{0}^{\ast}, x_{1}^{\ast},s)$ are linear
functionals of $\mbox{\boldmath $\psi$}^{\ast (1)} (x_{0}^{\ast},0)$. In
principle, this functional relationship can be inverted to give $%
\mbox{\boldmath $\psi$}^{\ast (2)}(x_{0}^{\ast},x_{1}^{\ast},s)= %
\mbox{\boldmath $\Psi$}^{\ast (2)} \left[ x_{0}^{\ast},x_{1}^{\ast},s| %
\mbox{\boldmath $\psi$}^{\ast (1)}\right]$. Use of this in Eq.\, (\ref{4.11}%
) provides the closed kinetic equation for $\mbox{\boldmath $\psi$}^{\ast
(1)}$. In practice it is a difficult many body problem to discover this
functional. However, its form is easily calculated at $s=0$. Equation (\ref%
{4.9}) gives 
\begin{equation}
\mbox{\boldmath $\Psi$}^{\ast (2)}\left[ x_{0}^{\ast},x_{1}^{\ast},s =0| %
\mbox{\boldmath $\psi$}^{\ast (1)} (0) \right]=f^{\ast (2)}(x_{0}^{\ast },
x_{1}^{\ast}) {\bf v}_{0}^{\ast } =f^{\ast (2)}(x_{0}^{\ast},x_{1}^{\ast}) %
\left[ f^{\ast (1)}(x_{0}^{\ast}) \right]^{-1}\mbox{\boldmath $\psi$}^{\ast
(1)} (x_{0}^{\ast},0),  \label{4.12}
\end{equation}
where $f^{\ast (1)}(x_{0}^{\ast}) $ and $f^{\ast
(2)}(x_{0}^{\ast},x_{1}^{\ast})$ and are the reduced distribution functions
associated with $\rho_{hcs}^{\ast}(\Gamma^{\ast })$. For fluids with elastic
collisions, the approximation 
\begin{equation}
\mbox{\boldmath $\Psi$}^{\ast (2)}\left[ x_{0}^{\ast},x_{1}^{\ast},s | %
\mbox{\boldmath $\psi$}^{\ast (1)}\right] \rightarrow \mbox{\boldmath $\Psi$}%
^{\ast (2)}\left[ x_{0}^{\ast},x_{1}^{\ast},0 | \mbox{\boldmath $\psi$}%
^{\ast (1)}(s) \right]  \label{4.13}
\end{equation}%
is accurate over a wide range of low to moderate densities. The same
approximation in the current case leads to the kinetic equation 
\begin{eqnarray}
\left( \partial _{s}+{\bf v}_{0}^{\ast} \cdot {\bf \nabla}^{\ast}_{0}
\right) \mbox{\boldmath $\psi$}^{\ast (1)}(x_{0}^{\ast},s) &+& \frac{%
\zeta^{\ast}}{2} \frac{\partial}{\partial {\bf v}_{0}^{\ast}} \cdot \left[ 
{\bf v}_{0}^{\ast} \mbox{\boldmath $\psi$}^{\ast (1)} (x_{0}^{\ast},s)\right]
\nonumber \\
&=&\int dx^{\ast}_{1}\,\overline{T}^{\ast}(0,1)f^{\ast (2)} (x_{0}^{\ast},
x_{1}^{\ast}) \left[ f^{\ast (1)} (x_{0}^{\ast}) \right]^{-1}%
\mbox{\boldmath
$\psi$}^{\ast (1)}(x^{\ast}_{0},s).  \label{4.14}
\end{eqnarray}
This equation is exact at short times by construction and its use for longer
times can be interpreted as a Markovian approximation \cite{LPyS69}. For the
VACF only the spatial integral of $\mbox{\boldmath $\psi$}^{\ast
(1)}(x^{\ast}_{0},s)$ is required, so the final representation of Eq.\ (\ref%
{4.10}) becomes 
\begin{equation}  \label{4.15}
C_{vv}^{\ast }(s)=\frac{2}{d\phi _{hcs}}\int d{\bf v}_{0}^{\ast } {\bf v}%
_{0}^{\ast} \cdot \mbox{\boldmath $\psi$}^{\ast}({\bf v}_{0}^{\ast},s),
\end{equation}
where 
\begin{equation}  \label{4.15a}
\mbox{\boldmath $\psi$}^{\ast}({\bf v}^{\ast}_{0},s)= \int d{\bf q}%
^{\ast}_{0}\, \mbox{\boldmath $\psi$}^{\ast (1)}(x^{\ast}_{0},s)
\end{equation}%
obeys the kinetic equation 
\begin{equation}
\partial _{s}\mbox{\boldmath $\psi$}^{\ast}({\bf v}^{\ast}_{0},s)+ \frac{%
\zeta ^{\ast }}{2} \frac{\partial}{\partial {\bf v}_{0}^{\ast}} \cdot \left[ 
{\bf v}^{\ast}_{0} \mbox{\boldmath $\psi$ }^{\ast}({\bf v}^{\ast}_{0},s) %
\right] ={\cal I}\mbox{\boldmath $\psi$}^{\ast}({\bf v}^{\ast}_{0},s).
\label{4.16}
\end{equation}
The collision operator ${\cal I}$ is defined as 
\begin{equation}
{\cal I}\mbox{\boldmath $\psi$}^{\ast}({\bf v}^{\ast}_{0},s) \equiv \int
dx^{\ast}_{1}\,\overline{T}^{\ast}(0,1)f^{\ast (2)} ( x_{0}^{\ast
},x_{1}^{\ast }) \left[ f^{\ast (1)}(x_{0}^{\ast}) \right]^{-1} %
\mbox{\boldmath $\psi$}({\bf v}^{\ast}_{0},s).  \label{4.17}
\end{equation}%
The kinetic equation (\ref{4.16}) has to be solved with the initial
condition 
\begin{equation}
\mbox{\boldmath $\psi$}^{\ast}({\bf v}^{\ast}_{0},0)=f^{\ast (1)} ({\bf v}%
_{0}^{\ast} ) {\bf v}_{0}^{\ast },  \label{4.16a}
\end{equation}%
as follows directly from Eq.\ (\ref{4.9}). To obtain Eq.\ (\ref{4.16}) from
Eq.\ (\ref{4.14}), use has been made of the fact that the HCS distributions
are spatially homogeneous. For $\alpha =1$ the linear collision operator $%
{\cal I}$ is non-negative, and the correlation function is found to decay on
a time scale of the order of $s \sim \omega _{1}^{\ast -1},$ the initial
rate of decay. Although not proven, it is reasonable to assume that the
spectrum of ${\cal I}$ is qualitatively similar for $\alpha <1$. Since the
kinetic equation is exact at $s\rightarrow 0$, the leading term in a
cumulant expansion of Eq.\ (\ref{4.9}) also is exact and agrees with Eq.\ (%
\ref{4.2}). More generally, this approximate kinetic equation gives
contributions to all higher terms in the cumulant expansion. However, for $%
\alpha =1$ these corrections are of the order of a few percent except when
the size or mass ratio of fluid and impurity particles differs greatly from
one.

The diffusion constant is obtained by integrating (\ref{4.10}) to get%
\begin{equation}
D^{\ast }=-\frac{2}{d\phi _{hcs}}\int d{\bf v}_{0}^{\ast }{\bf v}_{0}^{\ast
} \cdot {\bf X}\left( {\bf v}_{0}^{\ast }\right),  \label{4.21}
\end{equation}%
where ${\bf X}\left( {\bf v}_{0}^{\ast }\right) $ is the solution to the
integral equation%
\begin{equation}
{\cal I}X_{i}\left( {\bf v}_{0}^{\ast }\right) -\frac{\zeta ^{\ast }}{2} 
\frac{\partial}{\partial {\bf v}_{0}^{\ast}} \cdot \left[ {\bf v}%
^{\ast}_{0}X_{i}\left( {\bf v}^{\ast}_{0} \right) \right] =f^{\ast
(1)}\left( {\bf v}_{0}^{\ast }\right) v_{0i}^{\ast}.  \label{4.22}
\end{equation}%
The only (left) eigenfunction with vanishing eigenvalue of the operator
defined by the left hand side of the above equation is $1$. The right side
is orthogonal to $1$ so the Fredholm alternative (solubility condition) is
satisfied and a solution to this equation exists. It is shown in Appendix %
\ref{ap2} that the diffusion coefficient given by Eq.\ (\ref{4.21}) is the
same as that obtained from the Chapman-Enskog solution to the Enskog kinetic
equation, if velocity correlations are neglected for $f^{\ast
(2)}\left(x_{0}^{\ast },x_{1}^{\ast }\right)$ in the definition of ${\cal I}$%
, Eq.\ (\ref{4.17}). Recent computer simulations confirm that such neglect
is a good approximation \cite{SyM01}. Finally, if Eq.\ (\ref{4.22}) is
solved as an expansion in Sonine polynomials (the usual method for elastic
collisions), the first approximation yields again the leading cumulant
approximation. Thus the cumulant approximation, linear kinetic theory, and
Chapman-Enskog solution to the Enskog kinetic equation are all the same
modulo small differences due to velocity correlations and higher order
Sonine polynomials.

\section{Discussion}

\label{sec5} In this paper, it has been shown that standard linear response
theory can be extended in a natural way to describe diffusion in a system of
inelastic hard spheres in the homogeneous cooling state. The response
functions of the system are given in terms of stationary state averages
corresponding to a an effective dimensionless dynamics. In particular, the
dimensionless time scale is related to the average number of collisions per
particle taking place in the system, and it is the physically relevant one
to analyze the aging to a hydrodynamic stage. Similar Einstein and
Green-Kubo expressions have been obtained for the shear viscosity \cite%
{GyN99,DyB}, which is essentially a diffusion process like that considered
here. Response to an external force has been studied to determine
expressions for the mobility \cite{DyG01}. Transport coefficients associated
with longitudinal hydrodynamic modes (e.g. thermal conductivity) pose
special problems and will be discussed elsewhere. Beyond their formal
interest and utility for approximate analysis, the results derived here
enable a systematic nonperturbative study of transport processes in granular
fluids by means of molecular dynamics simulation of the response functions,
just as for fluids with elastic collisions.

It is interesting to consider the particular case of self-diffusion, i.e.
when the impurity particle is mechanically equivalent to the fluid
particles, for which previous analysis have been carried out. The expression
for the self-diffusion coefficient in the first cumulant approximation is
obtained by considering the limit $\phi_{hcs} =1$, $h=(1+\alpha )/4$, and $%
\alpha _{0}=\alpha $ in Eq.\ (\ref{4.7}). This yields 
\begin{equation}
D^{\ast }=\frac{d\Gamma \left( d/2\right) }{(1+\alpha )^{2}\chi 2^{1/2}\pi ^{%
\frac{d-1}{2}}}\, .  \label{5.1}
\end{equation}%
As expected, this result agrees with the expression derived in Ref. \cite%
{BRCyG00} from a Chapman-Enskog solution to the Enskog-Lorentz equation in a
first order polynomial expansion. A previous approach to self-diffusion in
the HCS, also in the context of linear response theory, has been developed
by Billiantov and P\"{o}schel \cite{ByP00}. Their result differs from Eq.\ (%
\ref{5.1}) in a factor of $(1+\alpha )/2$. Quite peculiarly, the same result
had been derived before by Hsiau and Hunt \cite{HyH93} and Savage and Dai %
\cite{SyD93}, independently, from approximate solutions of the
Enskog-Lorentz equation. The discrepancy between the result in Ref. \cite%
{ByP00} and the one reported here is not due to a different degree of
approximation, but has a fundamental physical origin. Brillantov and P\"{o}%
schel assume that the velocity autocorrelation function of the tagged
particle has the form 
\begin{equation}
\left[ C_{vv}(t,t^{\prime })\right] _{BP}=v(t)^{2}e^{-\frac{t-t^{\prime }}{%
\tau _{B}(t^{\prime })}},  \label{5.2}
\end{equation}%
with the relaxation time $\tau _{B}$ being inversely proportional to the
initial slope, in the actual time scale $t$, of the VACF. This is to be
contrasted with our analysis, based in the relevance of the time scale scale
defined by the number of collisions, 
\begin{equation}
C_{vv}(t,t^{\prime })=v(t)v(t^{\prime })\exp \left[ -\omega _{1}^{\ast
}\int_{t^{\prime }}^{t}d\tau \frac{v(\tau )}{\ell }\right],  \label{5.3}
\end{equation}%
with $\omega _{1}^{\ast }$ determined from the initial slope of the VACF in
the dimensionless scale defined by Eq.\ (\ref{2.22}). Making clear the
crucial role played by this latter time scale for the study of response
functions is one of the main goals in this paper.

The quality of the simple approximations given here is studied in the
following paper \cite{LDyB01} by comparison with molecular dynamics
simulations for both the Einstein and Green-Kubo forms. Only the case of
self-diffusion is considered. It is found that the agreement is very good
for low densities and all degrees of dissipation, but there are large
discrepancies at high density and large dissipation. The possible reasons
for the discrepancies are discussed there.

\section{Acknowledgments}

The research of JWD was supported by National Science Foundation grant PHY
9722133. JJB acknowledges partial support from the Direcci\'{o}n General de
Investigaci\'{o}n Cient\'{\i}fica y T\'{e}cnica (Spain) through Grant No.
PB98-1124. JFL acknowledges support from the Universit\'{e} Libre de
Bruxelles.

\appendix

\section{Cooling rates in the HCS}

\label{ap2} In this Appendix, the cooling rates for the fluid and the
impurity particle, as well as first cumulant for the VACF, are calculated
from the Liouville dynamics of the system. For simplicity, a local
equilibrium ensemble approximation, known to lead to accurate results \cite%
{GyD99}, will be considered. The dimensionless cooling rates for the fluid
and impurity particle are defined as 
\begin{equation}  \label{a1}
\zeta ^{\ast }\equiv -\frac{\ell}{v(t)} \frac{\partial \ln T}{\partial t}= -%
\frac{2}{d} \langle L^{\ast }{\bf v}_{1}^{\ast 2} \rangle^{*},
\end{equation}
\begin{equation}  \label{a2}
\zeta^{\ast}_{0} \equiv -\frac{\ell}{v(t)} \frac{\partial \ln T_{0}}{%
\partial t} =- \frac{2}{d \phi_{hcs}} \langle L^{\ast }{\bf v}_{0}^{\ast 2}
\rangle^{\ast},
\end{equation}
respectively. For the derivation of the last equalities, use has been made
of the property that in the HCS, for any dynamical variable $A(\Gamma)$
having the scaling property (\ref{2.30}) it is 
\begin{equation}  \label{a3}
\langle LA;t \rangle =\frac{ v^{a+1}(t)}{\ell}\, \langle L^{\ast} A (\{\ell 
{\bf q}^{\ast}_{i},{\bf v}^{\ast}_{i}\}) \rangle^{\ast}.
\end{equation}
The first cumulant for the VACF is given by Eq.\ (\ref{4.3}) or,
equivalently, 
\begin{equation}  \label{a4}
\omega _{1}^{\ast }= -\frac{\zeta^{\ast}}{2} -\frac{2}{d\phi _{hcs}} \langle
\left( L^{\ast }{\bf v}_{0}^{\ast }\right) \cdot {\bf v}_{0}^{\ast
}\rangle^{\ast}.
\end{equation}
In terms of the dimensionless distribution functions 
\begin{equation}  \label{a5}
f^{\ast (2)}(x_{0}^{\ast},x_{1}^{\ast}) =N \int dx^{\ast}_{2} dx^{\ast}_{3}
\ldots dx^{\ast}_{N} \rho^{*}_{hcs}(\Gamma^{\ast}),
\end{equation}
and 
\begin{equation}  \label{a6}
f^{\ast (2)}(x_{1}^{\ast},x_{2}^{\ast}) =N^{2} \int dx^{\ast}_{0}
dx^{\ast}_{3} \ldots dx^{\ast}_{N} \rho^{*}_{hcs}(\Gamma^{\ast}),
\end{equation}
the above quantities can be written as 
\begin{equation}  \label{a7}
\zeta^{\ast}= -\frac{2}{dN}\int dx^{\ast}_{1}dx^{\ast}_{2}f^{\ast (2)}\left(
x^{\ast}_{1},x^{\ast}_{2}\right) T^{\ast }(1,2){\bf v}_{1}^{\ast 2},
\end{equation}
\begin{equation}  \label{a8}
\zeta _{0}^{\ast }=- \frac{2}{d \phi_{hcs}} \int dx^{\ast}_{0} dx^{\ast}_{1}
f^{\ast (2)}\left( x^{\ast}_{0},x^{\ast}_{1}\right) T^{\ast }(0,1) {\bf v}%
_{0}^{\ast 2},
\end{equation}
\begin{equation}  \label{a9}
\omega _{1}^{\ast }=-\frac{\zeta ^{\ast }}{2}-\frac{2}{d\phi _{hcs}} \int
dx^{\ast}_{0}dx^{\ast}_{1}f^{\ast (2)}\left( x^{\ast}_{0},x^{\ast}_{1}
\right) {\bf v}_{0}^{\ast}\cdot \left[ T^{\ast }(0,1){\bf v}_{0}^{\ast } %
\right].
\end{equation}
The action of the binary collision operators can be performed for further
simplification, leading to 
\begin{equation}  \label{a10}
\zeta^{\ast}=\frac{1}{2dn^{\ast }}\left( 1-\alpha ^{2}\right) \sigma^{\ast
d-1} \int d{\bf v}_{1}^{\ast }\int d{\bf v}_{2}^{\ast }\int d\Omega f^{\ast
(2)} ({\bf v}_{1}^{\ast },{\bf v}_{2}^{\ast },{\bf q}^{\ast}_{12}= %
\mbox{\boldmath $\sigma$}^{\ast}) \Theta (-{\bf g}_{12}^{\ast }\cdot 
\widehat{\mbox{\boldmath $\sigma$}})|{\bf g}_{12}^{\ast }\cdot \widehat{%
\mbox{\boldmath $\sigma$}}|^{3},
\end{equation}
\begin{eqnarray}  \label{a11}
\zeta _{0}^{\ast} &=&-\frac{4h}{d\phi _{hcs}} \overline{\sigma}^{\ast d-1}
V^{\ast} \int d{\bf v}_{0}^{\ast } \int d{\bf v}_{1}^{\ast} \int d\Omega
f_{0}^{\ast (2)}({\bf v}_{0}^{\ast },{\bf v}_{1}^{\ast }, {\bf q}%
^{\ast}_{01}= \overline{\mbox{\boldmath $\sigma$}} ^{\ast})\Theta (-{\bf g}%
_{01}^{\ast }\cdot \widehat{\mbox{\boldmath $\sigma$}}) |{\bf g}_{01}^{\ast
}\cdot \widehat{\mbox{\boldmath $\sigma$}}|^{2}  \nonumber \\
&&\times \left[2{\bf G}_{01}^{\ast }\cdot \widehat{\mbox{\boldmath $\sigma$}}
+\frac{m}{m+m_{0}} \left( 1-\alpha _{0}\right) \widehat{%
\mbox{\boldmath
$\sigma$}} \cdot {\bf g}_{01}^{\ast } \right],
\end{eqnarray}
\begin{eqnarray}  \label{a12}
\omega _{1}^{\ast } &=&-\frac{\zeta^{\ast}}{2}-\frac{4h}{d\phi_{hcs}} 
\overline{\sigma}^{\ast d-1} V^{*} \int d{\bf v}_{0}^{\ast } \int d{\bf v}%
_{1}^{\ast } \int d\Omega f^{\ast (2)}({\bf v}_{0}^{\ast}, {\bf v}%
_{1}^{\ast},{\bf q}_{01}^{\ast}= \overline{\mbox{\boldmath $\sigma$}} )
\Theta (-{\bf g}_{01}^{\ast }\cdot \widehat{\mbox{\boldmath $\sigma$}}) |%
{\bf g}_{01}^{\ast} \cdot \widehat{\mbox{\boldmath $\sigma$}}|^{2}  \nonumber
\\
&&\times \left( {\bf G}_{01}^{\ast} \cdot \widehat{\mbox{\boldmath $\sigma$}}
+\frac{m}{m_{0}+m} {\bf g}_{01} \cdot \widehat{\mbox{\boldmath $\sigma$}}
\right),
\end{eqnarray}
where $h$ is defined below Eq.\ (\ref{3.17a}), $V^{\ast}$ is the reduced
volume or surface of the system, and ${\bf G}^{\ast}_{01}$ is the reduced
velocity of the center of mass for particles $0$ and $1$, i.e. 
\begin{equation}  \label{a12a}
{\bf G}^{\ast}_{01}=\frac{m_{0}{\bf v}^{\ast}_{0} +m {\bf v}^{\ast}_{1}}{%
m_{0}+m}.
\end{equation}
The above results are still exact. To compute the integrals two
approximations are introduced. First, the velocity correlations in $%
f^{(2)\ast }({\bf v}_{1}^{\ast },{\bf v}_{2}^{\ast },{\bf q}_{12}^{\ast}= %
\mbox{\boldmath $\sigma$}^{\ast})$ and $f^{\ast (2)}\left( {\bf v}_{0}^{\ast
},{\bf v}_{1}^{\ast },{\bf q}_{01}^{\ast}= \overline{%
\mbox{\boldmath
$\sigma$}}^{\ast}\right) $ are neglected. Note that this approximation is
required only for particles at contact and for the precollision hemisphere.
Significant velocity correlations exist on the opposite hemisphere and at
larger separation of the particles, and no restriction is placed on these
configurations. The second approximation is to represent the one particle
distributions by Maxwellians. This is known to be a good approximation
except for conditions of large mechanical differences between the fluid and
impurity particles. These two approximations are equivalent to write 
\begin{equation}  \label{a13}
f^{\ast (2)}\left( {\bf v}_{0}^{\ast },{\bf v}_{1}^{\ast },{\bf q}%
_{01}^{\ast} =\overline{\mbox{\boldmath $\sigma$}}\right) = n^{\ast} \chi
_{0} (\overline{\sigma}) \varphi_{0}^{\ast}({\bf v}_{0}^{\ast})
\varphi^{\ast} ({\bf v}_{1}^{\ast}),
\end{equation}
\begin{equation}  \label{a14}
f^{\ast (2)}\left( {\bf v}_{1}^{\ast },{\bf v}_{2}^{\ast },{\bf q}%
_{12}^{\ast} =\mbox{\boldmath $\sigma$}\right) = n^{\ast 2} \chi (\sigma)
\varphi^{\ast}({\bf v}_{1}^{\ast}) \varphi^{\ast} ({\bf v}_{2}^{\ast}),
\end{equation}
where $\chi_{0}(\overline{\sigma})$ and $\chi (\sigma)$ are the
fluid-impurity and and fluid-fluid pair correlation functions for particles
at contact, and 
\begin{equation}  \label{a.15}
\varphi_{0}^{\ast}({\bf v}_{0}^{\ast})= \frac{1}{\left( \phi_{hcs} \pi
\right)^{d/2}} e^{-\frac{v_{0}^{\ast 2}}{\phi_{hcs}}},
\end{equation}
\begin{equation}  \label{a.16}
\varphi^{\ast}({\bf v}_{i}^{\ast})=\frac{1}{\pi^{d/2}} e^{-v_{1}^{\ast 2}}.
\end{equation}
Now the integrations in Eqs.\ (\ref{a10})-(\ref{a12}) can be performed. The
calculations are straightforward but lengthy, and will be not reproduce
here. The results are given by Eqs. (\ref{3.17})-(\ref{3.17b}), and (\ref%
{4.6}).

\section{Enskog Kinetic Theory}

\label{ap3}

In the main text sections, linear response has been applied to describe
diffusion directly from the Liouville dynamics. It is useful to see that
equivalent results, in appropriate approximations, follow from kinetic
theory as well. This is illustrated in this Appendix using the Enskog
kinetic theory, expected to be valid over a wide range of densities and
restitution coefficients. To review its origin and applicability to granular
fluids, the first hierarchy equations of (\ref{2.14}) for the impurity
particle reduced distribution function $f_{0}^{(1)}({\bf q}_{0},{\bf v}%
_{0},t)$ and one particle fluid reduced distribution function $f^{(1)}({\bf q%
}_{1},{\bf v}_{1},t)$, are written explicitly, 
\begin{eqnarray}
\left( \partial _{t}+{\bf v}_{0}\cdot \nabla_{0} \right) f_{0}^{(1)}({\bf q}%
_{0},{\bf v}_{0},t) &=&\overline{\sigma }^{d-1} \int d{\bf q}_{1} \int d{\bf %
v}_{1} \int d\Omega\, \Theta (-{\bf g}_{01} \cdot \widehat{%
\mbox{\boldmath
$\sigma$}}) |{\bf g}_{01}{\bf \cdot \widehat{\mbox{\boldmath  $\sigma $}}|} 
\nonumber \\
&&\times \left[ \alpha _{0}^{-2}\delta ({\bf q}_{01} -\overline{%
\mbox{\boldmath $\sigma $}})b_{01}^{-1}-\delta ({\bf q}_{01}+ \overline{%
\mbox{\boldmath $\sigma $}})\right] f^{(2)}({\bf q}_{0},{\bf v}_{0},{\bf q}%
_{1},{\bf v}_{1},t),  \label{b1}
\end{eqnarray}%
\begin{eqnarray}
\left( \partial _{t}+{\bf v}_{1}\cdot \nabla_{1} \right) f^{(1)}({\bf q}%
_{1}, {\bf v}_{1},t) &=&\sigma^{d-1} \int d{\bf q}_{2} \int d{\bf v}_{2}
\int d\Omega\, \Theta (-{\bf g}_{12} \cdot \widehat{%
\mbox{\boldmath $\sigma
$}}) |{\bf g}_{12} \cdot \widehat{\mbox{\boldmath $\sigma $}}|  \nonumber \\
&&\times \left[ \alpha ^{-2}\delta ({\bf q}_{12}- \mbox{\boldmath $\sigma $}%
)b_{12}^{-1}-\delta ({\bf q}_{12}+ \mbox{\boldmath $\sigma $}) \right]
f^{(2)}({\bf q}_{1},{\bf v}_{1},{\bf q}_{2},{\bf v}_{2},t),  \label{b2}
\end{eqnarray}%
were $f^{(2)}({\bf q}_{0},{\bf v}_{0},{\bf q}_{1},{\bf v}_{1},t)$ and $%
f^{(2)}({\bf q}_{1},{\bf v}_{1},{\bf q}_{2},{\bf v}_{2},t)$ are the two
particle reduced distribution functions for the impurity and one fluid
particle, and for two fluid particles, respectively. Their definitions are
given by Eqs. (\ref{2.15}) and (\ref{2.19a}). A closure of the hierarchy is
obtained by replacing in the above 
\begin{equation}
f^{(2)}({\bf q}_{0},{\bf v}_{0},{\bf q}_{1},{\bf v}_{1},t)\rightarrow
\chi_{0}({\bf q}_{0},{\bf q}_{1};t)f_{0}^{(1)} ({\bf q}_{0},{\bf v}_{0},t)
f^{(1)}({\bf q}_{1},{\bf v}_{1},t),  \label{b3}
\end{equation}
\begin{equation}
f^{(2)}({\bf q}_{1},{\bf v}_{1},{\bf q}_{2},{\bf v}_{2},t)\rightarrow \chi (%
{\bf q}_{1},{\bf q}_{2};t)f^{(1)}({\bf q}_{1},{\bf v}_{1},t) f^{(1)}({\bf q}%
_{2},{\bf v}_{2},t).  \label{b4}
\end{equation}%
The approximation is a generalization of that in (\ref{4.13}), and also of
that in Eqs. (\ref{a13}) and (\ref{a14}). It can be understood in two
different ways. It is the exact Markovian limit if the initial distribution
functions have the forms (\ref{b3}) and (\ref{b4}) on the precollisional
hemisphere (exact at short times). It also follows if velocity correlations
on this hemisphere are destroyed between collisions (Boltzmann's argument).
In any case, it is known to provide a good description of the hard sphere
fluid over a wide range of densities and times for elastic collisions, and
this validity appears to hold as well for inelastic collisions. As in the
elastic collisions case, the functions $\chi _{0}({\bf q}_{0},{\bf q}_{1};t)$
and $\chi ({\bf q}_{1},{\bf q}_{2};t)$ are taken to be the equilibrium
configurational pair correlation functions as functionals of the
nonequilibrium density.

The Enskog approximation converts Eqs.\ (\ref{b3}) and (\ref{b4}) into a
pair of kinetic equations. The kinetic equation for the fluid is autonomous
while that for the impurity distribution is a functional of $f^{(1)}({\bf q}%
_{1},{\bf v}_{1},t),$ 
\begin{equation}
\left( \partial _{t}+{\bf v}_{1}\cdot \nabla_{1} \right) f({\bf q}_{1},{\bf v%
}_{1},t)=J_{E}\left[ {\bf q}_{1},{\bf v}_{1}|f(t) \right],  \label{b5}
\end{equation}
\begin{equation}
\left( \partial _{t}+{\bf v}_{0}\cdot \nabla_{0} \right) f_{0}({\bf q}_{0},%
{\bf v}_{0},t)= J_{EL}\left[{\bf q}_{0},{\bf v}_{0}|f_{0}(t),f(t)\right].
\label{b6}
\end{equation}%
Here and below the superscript on the single particle distributions will be
suppressed for simplicity of notation. The Enskog and Enskog-Lorentz
collision operators are given by 
\begin{eqnarray}
J_{E}[{\bf q}_{1},{\bf v}_{1}|f(t)] &\equiv &\sigma ^{d-1}\int d{\bf v}%
_{2}\int d\Omega\, \Theta (-{\bf g}_{12} \cdot \widehat{%
\mbox{\boldmath 
$\sigma$}}) |{\bf g}_{12} \cdot \widehat{\mbox{\boldmath  $\sigma$}}| 
\nonumber \\
&& \times \left\{ \alpha ^{-2} \chi \left[ {\bf q}_{1},{\bf q}_{1}- %
\mbox{\boldmath $\sigma$} | n(t) \right] b_{12}^{-1} f({\bf q}_{1},{\bf v}%
_{1},t)f({\bf q}_{1}-\mbox{\boldmath $\sigma$}, {\bf v}_{2},t) \right. 
\nonumber \\
&&\left. -\chi \left[ {\bf q}_{1},{\bf q}_{1}+ \mbox{\boldmath $\sigma$}|
n(t) \right] f({\bf q}_{1},{\bf v}_{1},t) f({\bf q}_{1}+ 
\mbox{\boldmath
$\sigma$} ,{\bf v}_{2},t)\right\} ,  \label{b8}
\end{eqnarray}%
\begin{eqnarray}
J_{EL}[{\bf q}_{0},{\bf v}_{0}|f_{0}(t),f(t)] &\equiv &\overline{\sigma }%
^{d-1} \int d{\bf v}_{1} \int d\Omega\, \Theta (-{\bf g}_{01} \cdot \widehat{%
\mbox{\boldmath $\sigma$}}) |{\bf g}_{01} \cdot \widehat{%
\mbox{\boldmath
$\sigma$}}|  \nonumber \\
& & \times \left\{\alpha _{0}^{-2} \chi_{0} \left[ {\bf q}_{0},{\bf q}_{0} -%
\overline{\mbox{\boldmath $\sigma$}} |n(t) \right] b_{01}^{-1} f_{0} ({\bf q}%
_{0},{\bf v}_{0},t) f({\bf q}_{0}- \overline{\mbox{\boldmath $\sigma$}},{\bf %
v}_{1},t) \right.  \nonumber \\
& & \left. -\chi _{0}\left[ {\bf q}_{0},{\bf q}_{0}+ \overline{%
\mbox{\boldmath $\sigma$}} |n(t) \right] f_{0} ({\bf q}_{0}, {\bf v}_{0},t)f(%
{\bf q}_{0}+\overline{\mbox{\boldmath $\sigma$}},{\bf v}_{1}, t) \right\} ,
\label{b9}
\end{eqnarray}%
respectively. The HCS is a special case for which Eqs.\ (\ref{b5}) and (\ref%
{b6}) reduce to 
\begin{equation}
\partial _{t}f_{hcs}=J_{E}\left[ {\bf v}_{1}|f_{hcs}(t) \right],  \label{b10}
\end{equation}%
\begin{equation}
\partial _{t}f_{0,hcs}=J_{EL}\left[ {\bf v}_{0}|f_{0,hcs}(t),f_{hcs}(t)%
\right], \;  \label{b10a}
\end{equation}%
with the collision terms 
\begin{equation}
J_{E}[{\bf v}_{1}|f_{hcs}(t)] \equiv \sigma ^{d-1}\chi \int d{\bf v}_{2}
\int d\Omega\, \Theta (-{\bf g}_{12} \cdot \widehat{\mbox{\boldmath $\sigma$}%
}) |{\bf g}_{12} \cdot \widehat{\mbox{\boldmath $\sigma$}} \left[ \left(
\alpha^{-2} b_{12}^{-1}-1 \right) f_{hcs}({\bf v}_{1},t,)f_{hcs}({\bf v}%
_{2},t,) \right] ,  \label{b11}
\end{equation}
\begin{eqnarray}
J_{EL}[{\bf v}_{0}|f_{0,hcs}(t),f_{hcs}(t)] &\equiv& \overline{\sigma }%
^{d-1}\chi_{0} \int d{\bf v}_{1}\int d \Omega\,\Theta (- {\bf g}_{01} \cdot 
\widehat{\mbox{\boldmath $\sigma$}}) | {\bf g}_{01} \cdot \widehat{%
\mbox{\boldmath $\sigma$}}|  \nonumber \\
&& \times \left( \alpha_{0}^{-2}b_{01}^{-1}-1 \right) f_{0,hcs}({\bf v}%
_{0},t)f_{hcs}({\bf v}_{1},t).  \label{b12}
\end{eqnarray}
Use has been made of the fact that $\chi \left[ {\bf q}_{1},{\bf q}_{1}- %
\mbox{\boldmath $\sigma$}|n \right]\equiv\chi $ and $\chi _{0}\left[ {\bf q}%
_{0},{\bf q}_{0}-\mbox{\boldmath $\sigma$} |n \right]\equiv \chi_{0}$ are
independent of space coordinates due to translational invariance.
Furthermore, the reduced distribution functions $f_{hcs}$ and $f_{0,hcs}$
have the scaling forms, inherited from (\ref{2.16}), 
\begin{eqnarray}
f_{hcs}({\bf v}_{1},t) &=&[\ell v( t) ]^{-d}f_{hcs}^{\ast }\left[ {\bf v}%
_{1}/v(t)\right],  \nonumber \\
f_{0,hcs}({\bf v}_{0},t) &=&[\ell v(t) ]^{-d}f_{0,hcs}^{\ast }\left[ {\bf v}%
_{0}/v(t)\right].  \label{b13}
\end{eqnarray}%
Substitution of these expressions into the kinetic equations leads to 
\begin{eqnarray}
\frac{1}{2}\zeta^{\ast} \frac{\partial}{\partial {\bf v}_{1}^{\ast}} \cdot
\left( {\bf v}_{1}f_{hcs}^{\ast}\right) &=& J_{E}^{\ast} \left[ {\bf v}_{1}|
f_{hcs}^{\ast }\right],  \nonumber \\
\frac{1}{2}\zeta^{\ast}_{0} \frac{\partial}{\partial {\bf v}_{0}^{\ast}}
\cdot \left( {\bf v}_{1}f_{0,hcs}^{\ast }\right) &=&J_{EL}^{\ast} \left[ 
{\bf v}_{0}|f_{0,hcs}^{\ast},f_{hcs}^{\ast }\right].  \label{b14}
\end{eqnarray}
where the scaled position and velocity variables defined below Eq.\ (\ref%
{2.26}) have been again introduced. The solution to these equations has been
discussed elsewhere \cite{GyD99b} and will not be considered further here.

Now consider more general spatially inhomogeneous states for the impurity
particle, with the fluid still in the HCS. The probability density to find
the impurity at a point ${\bf q}_{0}$ in terms of the reduced distribution
function is given by 
\begin{equation}
P({\bf q}_{0},t)=\int d{\bf v}_{0}f_{0}({\bf q}_{0},{\bf v}_{0},t).
\label{b15}
\end{equation}%
Integration of Eq.\ (\ref{b6}) over the velocity ${\bf v}_{0}$ gives the
conservation law for probability 
\begin{equation}
\partial _{t}P({\bf q}_{0},t)+{\bf \nabla }_{0}\cdot {\bf J}({\bf q}%
_{0},t)=0,  \label{b16}
\end{equation}%
\begin{equation}
{\bf J}({\bf q}_{0},t)=\int d{\bf v}_{0}{\bf v}_{0}f_{0}({\bf q}_{0},{\bf v}%
_{0},t)  \label{b16a}
\end{equation}%
which, of course, are the same as Eqs. (\ref{3.2}) and (\ref{3.3}). The
diffusion equation is obtained from a ``normal'' solution to the
Boltzmann-Lorentz equation, where all space and time dependence occurs
through $P({\bf q}_{0},t)$ and $T(t)$. The linear relationship (\ref{b15})
implies that such a solution has the form 
\begin{equation}
f_{0}({\bf q}_{0},{\bf v}_{0},t)=P({\bf q}_{0},t)h\left[ {\bf v}_{0}/v(t)%
\right] .  \label{b17}
\end{equation}%
The Chapman-Enskog method represents a normal solution to the kinetic
equation as an expansion in the gradients, 
\begin{equation}
f_{0}({\bf q}_{0},{\bf v}_{0},t)=f_{0}^{(0)}({\bf q}_{0},{\bf v}%
_{0},t)+\epsilon f_{0}^{(1)}({\bf q}_{0},{\bf v}_{0},t)+\ldots ,  \label{b18}
\end{equation}%
where $\epsilon $ is a formal parameter representing the order of the
spatial gradient. Similarly, the time derivative is obtained as an expansion
in the gradients via the conservation equation, 
\begin{equation}
\partial _{t}=\partial _{t}^{(0)}+\epsilon \partial _{t}^{(1)}+\ldots
\label{b19}
\end{equation}%
and the kinetic equation is written%
\begin{equation}
\left( \partial _{t}+\epsilon {\bf v}_{0}\cdot \nabla _{0}\right) f_{0}({\bf %
q}_{0},{\bf v}_{0},t)=J_{EL}\left[ {\bf q}_{0},{\bf v}%
_{0}|f_{0}(t),f_{hcs}(t)\right] .  \label{b20}
\end{equation}%
Substitution of Eqs.\ (\ref{b18}) and (\ref{b19}) into Eq.\ (\ref{b20})
gives to zeroth order in the spatial gradients%
\begin{equation}
\partial _{t}^{(0)}f_{0}^{(0)}=J_{EL}\left[ {\bf q}_{0},{\bf v}%
_{0}|f_{0}^{(0)}(t),f_{hcs}(t)\right] ,  \label{b21}
\end{equation}%
which has the solution%
\begin{equation}
f_{0}^{(0)}({\bf q}_{0},{\bf v}_{0},t)=V P({\bf q}_{0},t)f_{hcs}({\bf v}%
_{0},t).  \label{b22}
\end{equation}%
This gives ${\bf J}^{(0)}=0$ and, consequently, $\partial _{t}^{(1)}P=0$.
The first order correction $f_{0}^{(1)}$ is determined from%
\begin{equation}
\partial _{t}^{(0)}f_{0}^{(1)}+\left( \partial _{t}^{(1)}+{\bf v}_{0}\cdot
\nabla _{0}\right) f_{0}^{(0)}=J_{EL}\left[ {\bf q}_{0},{\bf v}%
_{0}|f_{0}^{(1)}(t),f_{hcs}(t)\right] .  \label{b23}
\end{equation}%
The contribution from $\partial _{t}^{(1)}$ vanishes, since it is
proportional to $\partial _{t}^{(1)}P({\bf q}_{0},t)=-{\bf \nabla }_{0}\cdot 
{\bf J}^{(0)}({\bf q}_{0},t)=0$. Also, the time derivative $\partial
_{t}^{(0)}$ can be expressed in terms of the cooling rate $\zeta $. Using
once again the dimensionless variables defined in the main text, Eq.\ (\ref%
{b23}) becomes 
\begin{equation}
{\cal I}_{EL}f_{0}^{\ast (1)}-\frac{\zeta ^{\ast }}{2}\frac{\partial }{%
\partial {\bf v}_{0}^{\ast }}\cdot \left( {\bf v}_{0}^{\ast }f_{0}^{\ast
(1)}\right) =f_{0,hcs}^{\ast }{\bf v}_{0}^{\ast }\cdot \nabla _{0}^{\ast
}P^{\ast }({\bf q}_{0}^{\ast },t),  \label{b24}
\end{equation}%
where ${\cal I}_{EL}$ is the linear Enskog-Lorentz collision operator, 
\begin{equation}
{\cal I}_{EL}f_{0}^{\ast (1)}\equiv \int d{\bf v}_{1}^{\ast }\int d\Omega
\,\Theta (-{\bf g}^{\ast}_{01}\cdot \widehat{\mbox{\boldmath $\sigma$}})| 
{\bf g}^{\ast}_{01}\cdot \widehat{\mbox{\boldmath $\sigma$}}|\left( \alpha
_{0}^{-2}b_{01}^{-1}-1\right) f^{\ast }({\bf v}_{1}^{\ast })f_{0}^{\ast (1)}(%
{\bf q}_{0}^{\ast },{\bf v}_{0}^{\ast },t),  \label{b25a}
\end{equation}%
that is the same as the collision operator ${\cal I}$ in Eq.\ (\ref{4.17})
if velocity correlations, associated to the two particle distribution, are
neglected in the latter. The solution to Eq.\thinspace\ (\ref{b24}) can be
written in the form 
\begin{equation}
f_{0}^{\ast (1)}({\bf v}_{0}^{\ast },{\bf q}^{\ast },t)={\bf X}\left( {\bf v}%
_{0}^{\ast }\right) \cdot \nabla _{0}^{\ast }P^{\ast }({\bf q}^{\ast },s),
\label{b25b}
\end{equation}%
where ${\bf X}({\bf v}_{0}^{\ast })$ is the solution to the integral
equation 
\begin{equation}
{\cal I}_{EL}X_{i}\left( {\bf v}_{0}^{\ast }\right) -\frac{\zeta ^{\ast }}{2}%
\frac{\partial }{\partial {\bf v}_{0}^{\ast }}\cdot \left[ {\bf v}_{0}^{\ast
}X_{i}({\bf v}_{0}^{\ast })\right] =f_{0,hcs}^{\ast }({v}_{0}^{\ast
})v_{0i}^{\ast }.  \label{b25}
\end{equation}%
Finally, the diffusion coefficient is identified from%
\begin{equation}
{\bf J}^{(1)}({\bf q}_{0},t)=\int d{\bf v}_{0}\,{\bf v}_{0}f_{0}^{(1)}({\bf q%
}_{0},{\bf v}_{0},t)=-D\nabla_{0} P({\bf q}_{0},t)  \label{b25c}
\end{equation}%
to get%
\begin{equation}
D^{\ast }=-\frac{2}{d\phi _{hcs}}\int d{\bf v}_{0}^{\ast }{\bf v}_{0}^{\ast
}\cdot {\bf X}({\bf v}_{0}^{\ast }).  \label{b26}
\end{equation}%
These are the same results as those from Sec. \ref{sec4}, Eqs. (\ref{4.21})
and (\ref{4.22}).

\end{document}